%Paper: cond-mat/9410100
%From: "Bruno Nachtergaele" <bxn@math.Princeton.EDU>
%Date: Wed, 26 Oct 94 16:16:38 EDT

%%%%%%%%%%%%%%%%%%%%%%%%%%%%%%%%%%%%%%%%%%%%%%%%%%%%%%%%%%%%%%%%%%
% This is a Plain TeX file with 3 figures appended as postscript %
% files. Simply TeXing the file will produce a copy without      %
% figures. In order to print the figures they have to be saved in%
% separate files, beginning with a line "%!PS-Adobe-2.0" and     %
% ending with a line "%%EOF". These files can then be sent to a  %
% postscript printer. In the present file the postscript files   %
% are preceded by a line containing - cut here - and the name of %
% the file, - fig1.ps - etc., which has to be erased.            %
%%    %
% Title: The Stability of the Peierls Instability                %
%        for Ring-Shaped Molecules                               %
% Authors: Elliott Lieb and Bruno Nachtergaele                   %
% Address: Department of Physics                                 %
%          P. O. Box 708                                         %
%          Princeton, NJ 08544-0708                              %
%          USA                                                   %
% E-mail: bxn@math.princeton.edu (Bruno Nachtergaele)            %
%         lieb@math.princeton.edu (Elliott Lieb)                 %
% Date created: 26 June 1994                                     %
%%%%%%%%%%%%%%%%%%%%%%%%%%%%%%%%%%%%%%%%%%%%%%%%%%%%%%%%%%%%%%%%%%
\magnification=\magstep1 \relax
\overfullrule=0pt
\hsize=17truecm \vsize=22 truecm \hoffset=0.0truecm \voffset=0.0truecm
\parskip=0pt
\baselineskip=12pt

\def\draftonly#1{\ifx\draft1{\rm \lbrack{\ttraggedright {#1}}
                    \rbrack\quad}                \else\fi}

\def\today{\ifcase\month\or
 January\or February\or March\or April\or May\or June\or
 July\or August\or September\or October\or November\or December\fi
 \space\number\day, \number\year}

\font\BF=cmbx10 scaled \magstep 3
 5
\font\eightrm=cmr8

\let\1\sp

%\headline={\eightrm Lieb and Nachtergaele, Stability of the Peierls
%Instability\hfill version of \today\hfill\number\pageno}

% THEOREMS  : allow items in proclaim
\def\lessblank{\parskip=5pt \abovedisplayskip=2pt
          \belowdisplayskip=2pt }
%\outer\def\iproclaim #1. { %plus5pt
%        \par \vskip10pt\noindent
%    {\bf #1.\ }\begingroup\interlinepenalty=250\lessblank\it}
\def\eproclaim{\par\endgroup\vskip10pt plus5pt
               \noindent}
\def\proof#1{\par\noindent {\bf Proof #1}\          % Use as "\proof:"
         \begingroup\lessblank\parindent=0pt}
\def\QED {\hfill {\bf Q.E.D.}\endgroup\break\smallskip}
\def\QEDnogroup{\hfill {\bf Q.E.D.} \break\smallskip}

\def\nl{\hfill\break}

\def\titem#1\par{\item{$\triangleright$} #1\par \vskip-8pt}

%\input xrf.tex %xrd.tex for draft
% *************************************
% * This is the file: xrf.tex
% ** Tue  04-19-1994  *****************
\catcode`@=11
%%%%%%%%%%%% CROSS REFERENCE PACKAGE %%%%%%%%%%%%%%%%%%%%%%%%
\def\ifundefined#1{\expandafter\ifx\csname
                        \expandafter\eat\string#1\endcsname\relax}
\def\atdef#1{\expandafter\def\csname #1\endcsname}
\def\atedef#1{\expandafter\edef\csname #1\endcsname}
\def\atname#1{\csname #1\endcsname}
\def\ifempty#1{\ifx\@mp#1\@mp}
\def\ifatundef#1#2#3{\expandafter\ifx\csname#1\endcsname\relax
                                  #2\else#3\fi}
\def\eat#1{}

%%%%%%%%%%%%%%%%%%%%%%%% CITATIONS %%%%%%%%%%%%%%%%%%%%%%%%%%%
\newcount\refno \refno=1
\def\labref #1 #2 #3\par{\atdef{R@#2}{#1}}
\def\lstref #1 #2 #3\par{\atedef{R@#2}{\number\refno}
                              \advance\refno by1}
\def\txtref #1 #2 #3\par{\atdef{R@#2}{\number\refno
      \global\atedef{R@#2}{\number\refno}\global\advance\refno by1}}
\def\doref  #1 #2 #3\par{{\refno=0
     \vbox {\everyref \item {\reflistitem{\atname{R@#2}}}
            {\d@more#3\more\@ut\par}\par}}\vskip\refskip }
\def\d@more #1\more#2\par
   {{#1\more}\ifx#2\@ut\else\d@more#2\par\fi}
\let\more\relax
\let\everyref\relax  % executed before every item
\newdimen\refskip  \refskip=\parskip
\let\REF\labref
\def\@cite #1,#2\@ver
   {\eachcite{#1}\ifx#2\@ut\else,\@cite#2\@ver\fi}
\def\citeform#1{\lbrack{\bf#1}\rbrack}
\def\cite#1{\citeform{\@cite#1,\@ut\@ver}}
\def\eachcite#1{\ifatundef{R@#1}{{\tt#1??}}{\atname{R@#1}}}
\def\defonereftag#1=#2,{\atdef{R@#1}{#2}}
\def\defreftags#1, {\ifx\relax#1\relax \let\next\relax \else
           \expandafter\defonereftag#1,\let\next\defreftags\fi\next }

%%%%%%%%%%%%%%%citation formats: %%%%%%%%%%%%%%%%%%%%%%%%%%
\newdimen\refskip  \refskip=\parskip
\def\@utfirst #1,#2\@ver
   {\author#1,\ifx#2\@ut\afteraut\else\@utsecond#2\@ver\fi}
\def\@utsecond #1,#2\@ver
   {\ifx#2\@ut\andone\author#1,\afterauts\else
      ,\author#1,\@utmore#2\@ver\fi}
\def\@utmore #1,#2\@ver
   {\ifx#2\@ut\and\author#1,\afterauts\else
      ,\author#1,\@utmore#2\@ver\fi}
\def\authors#1{\@utfirst#1,\@ut\@ver}
\def\citeform#1{{\bf\lbrack#1\rbrack}}
\let\everyref\relax            % executed before every item
\let\more\relax                % executed after every subitem
\let\reflistitem\citeform
\catcode`@=12
\def\Bref#1 "#2"#3\more{\authors{#1}:\ {\it #2}, #3\more}
\def\Gref#1 "#2"#3\more{\authors{#1}\ifempty{#2}\else:``#2''\fi,
                             #3\more}
\def\Jref#1 "#2"#3\more{\authors{#1}:``#2'', \Jn#3\more}
\def\inPr#1 "#2"#3\more{in: \authors{\eds#1}:``{\it #2}'', #3\more}
\def\Jn #1 @#2(#3)#4\more{{\it#1}\ {\bf#2}(#3)#4\more}
\def\author#1. #2,{#1.~#2}
%% use as "\sameauthor. {}"
\def\sameauthor#1{\leavevmode$\underline{\hbox to 25pt{}}$}
\def\and{, and}   \def\andone{ and}
%% use as    \noinitial. Wolfram Research
\def\noinitial#1{\ignorespaces}
\let\afteraut\relax
\let\afterauts\relax
\def\etal{\def\afteraut{, et.al.}\let\afterauts\afteraut
           \let\and,}
\def\eds{\def\afteraut{(ed.)}\def\afterauts{(eds.)}}
\catcode`@=11

%%%%%%%%%%%%%% EQUATION NUMBERS %%%%%%%%%%%%%%%%%%%%
\newcount\eqNo \eqNo=0
\def\lasteq{\secNo.\number\eqNo}
\def\deq#1(#2){{\ifempty{#1}\global\advance\eqNo by1
       \edef\n@@{\lasteq}\else\edef\n@@{#1}\fi
       \ifempty{#2}\else\global\atedef{E@#2}{\n@@}\fi\n@@}}
\def\eq#1(#2){\edef\n@@{#1}\ifempty{#2}\else
       \ifatundef{E@#2}{\global\atedef{E@#2}{#1}}%
                       {\edef\n@@{\atname{E@#2}}}\fi
       {\rm(\n@@)}}
\def\deqno#1(#2){\eqno(\deq#1(#2))}
\def\deqal#1(#2){(\deq#1(#2))}
\def\eqback#1{{(\advance\eqNo by -#1 \lasteq)}}

\def\eqgroup(#1){{\global\advance\eqNo by1
       \edef\n@@{\lasteq}\global\atedef{E@#1}{\n@@}}}

%%%%%%%%%%%%%%%%% PROCLAIM NUMBERS %%%%%%%%%%%%%%%%%%%%%%%%%
\outer\def\iproclaim#1/#2/#3. {\vskip10pt plus5pt \par\noindent
     {\bf\dpcl#1/#2/ #3.\ }\begingroup \interlinepenalty=1000
\lessblank\it}
\newcount\pcNo  \pcNo=0
\def\lastpc{\number\pcNo} % NOT by section

\def\dpcl#1/#2/{\ifempty{#1}\global\advance\pcNo by1
       \edef\n@@{\lastpc}\else\edef\n@@{#1}\fi
       \ifempty{#2}\else\global\atedef{P@#2}{\n@@}\fi\n@@}
\def\pcl#1/#2/{\edef\n@@{#1}%
       \ifempty{#2}\else
       \ifatundef{P@#2}{\global\atedef{P@#2}{#1}}%
                       {\edef\n@@{\atname{P@#2}}}\fi
       \n@@}
\def\Def#1/#2/{Definition~\pcl#1/#2/}
\def\Thm#1/#2/{Theorem~\pcl#1/#2/}
\def\Lem#1/#2/{Lemma~\pcl#1/#2/}
\def\Prp#1/#2/{Proposition~\pcl#1/#2/}
\def\Cor#1/#2/{Corollary~\pcl#1/#2/}
\def\Exa#1/#2/{Example~\pcl#1/#2/}

%%%%%%%%%%%%%%%% SECTIONS & TABLE OF CONTENTS %%%%%%%%%%%
\font\sectfont=cmbx10 %scaled \magstep1
\def\secNo{00}
\def\Beginsection#1#2{\vskip\z@ plus#1\penalty-250
  \vskip\z@ plus-#1\bigskip\vskip\parskip
  %\leftline
\bigskip\noindent{\bf#2}%\nobreak
                           \smallskip\noindent}
\def\bgsection#1. #2\par{\Beginsection{.3\vsize}{\sectfont#1.\ #2 }%
            \def\secNo{#1}\eqNo=0}
\def\bgssection#1. #2\par{\Beginsection{.3\vsize}{#1.\ #2 }%
            \def\secNo{#1}\eqNo=0}
\def\Acknow#1\par{\ifx\REF\doref
     \Beginsection{.3\vsize}{\sectfont Acknowledgements}%
#1\par
     \Beginsection{.3\vsize}{\sectfont References}\fi}
\catcode`@=12
%%%%%%%%%%%%%%%%%% GENERAL STUFF %%%%%%%%%%%%%%%%%%%%%%%%%%
% for AMS-classification
\def\class#1 #2*{{#1},}
\overfullrule=0pt

% BLACKBOARD BOLD
\def\idty{{\leavevmode{\rm 1\ifmmode\mkern -5.4mu\else
                                            \kern -.3em\fi I}}}
\def\Ibb #1{ {\rm I\ifmmode\mkern -3.6mu\else\kern -.2em\fi#1}}
\def\Ird{{\hbox{\kern2pt\vbox{\hrule height0pt depth.4pt width5.7pt
    \hbox{\kern-1pt\sevensy\char"36\kern2pt\char"36} \vskip-.2pt
    \hrule height.4pt depth0pt width6pt}}}}
\def\Irs{{\hbox{\kern2pt\vbox{\hrule height0pt depth.34pt width5pt
       \hbox{\kern-1pt\fivesy\char"36\kern1.6pt\char"36} \vskip -.1pt
       \hrule height .34 pt depth 0pt width 5.1 pt}}}}
\def\Ir{{\mathchoice{\Ird}{\Ird}{\Irs}{\Irs} }}
\def\ibbt #1{\leavevmode\hbox{\kern.3em\vrule
     height 1.5ex depth -.1ex width .2pt\kern-.3em\rm#1}}
\def\ibbs#1{\hbox{\kern.25em\vrule
     height 1ex depth -.1ex width .2pt
                   \kern-.25em$\scriptstyle\rm#1$}}
\def\ibbss#1{\hbox{\kern.22em\vrule
     height .7ex depth -.1ex width .2pt
                   \kern-.22em$\scriptscriptstyle\rm#1$}}
\def\ibb#1{{\mathchoice{\ibbt #1}{\ibbt #1}{\ibbs #1}{\ibbss #1}}}
 \def\Cx {{\ibb C}} \def\Rl {{\Ibb R}}

% OPERATORS

  % redefinition!
\def\Order{{\rm O}}
  % redefinition!

  % also in text
\def\tover#1#2{{\textstyle{#1\over #2}}}
\def\tr{\mathop{\rm tr}\nolimits}
\def\Tr{\mathop{\rm Tr}\nolimits}

% LETTERS
\def\phi{\varphi}            % redefinition!
\def\epsilon{\varepsilon}    % redefinition!

   \def\D{{\cal D}}
 \def\H{{\cal H}}

\def\up{\uparrow}
\def\down{\downarrow}

\def\Nup{N_\uparrow}
\def\Ndown{N_\downarrow}

\let\tr\tright
\def\tl{t^{(l)}}
\def\tm{t^{(m)}}
\def\btr{t^{(r)*}}
\def\btl{t^{(l)*}}

\def\cd{c^{\dagger}}
\def\c{c^{\mathstrut}}
\def\es{{\bf S}}

\def\ee{{\cal E}}

\def\hc{\hbox{ h.c. }}
\def\half{{1/2}}
\def\imply{\Rightarrow}

\let\REF\lstref %\lstref for numbers ; \labref for letter labels
\REF Hue Hue \Gref\par
\REF Sal Sal \Bref\par
\REF Lon Lon \Jref\par
\REF Jon Jon \Jref\par
\REF Hub Hub \Jref\par
\REF PP PP \Jref\par
\REF Pop Pop \Jref\par
\REF Lie4 Lie4 \Gref\par
\REF SSH SSH \Jref\par
\REF KH KH \Jref\par
\REF BCM BCM \Gref\par
\REF BJ BJ \Gref\par
\REF Pei Pei \Bref\par
\REF Fro Fro \Jref\par
\REF LHS LHS \Jref\par
\REF Lab Lab \Jref\par
\REF Oos1 Oos1 \Jref\par
\REF KL KL \Jref\par
\REF Imr Imr \Gref\par
\REF Wie Wie \Jref\par
\REF Lie3 Lie3 \Jref\par
\REF LS LS \Jref\par
\REF Lie2 Lie2 \Jref\par
\REF Oos2 Oos2 \Jref\par
\REF Gar Gar \Bref\par
\REF And And \Jref\par
\REF Chs Chs \Jref\par
\REF Pin Pin \Jref\par
\REF BP BP \Jref\par
\REF CF CF \Jref\par
\REF SKM SKM \Jref\par
\REF AN AN \Gref\par
\REF BC BC \Bref\par
\REF DM DM \Jref\par
\REF KSSH KSSH \Jref\par
\REF MGEW MGEW \Jref\par
\REF EWGMO EWGMO \Jref\par
\REF FK FK \Jref\par
\REF ML ML \Jref\par
\REF PVZ PVZ \Jref\par
\REF SG SG \Jref\par
\REF HM HM \Jref\par
\REF MR MR \Jref\par
\REF UMT UMT \Gref\par
\REF Maz Maz \Jref\par
\REF TH TH \Jref\par
\REF KL2 KL2 \Jref\par
\REF BS BS \Jref\par
\REF LM LM \Gref\par
\REF Lie1 Lie1 \Jref\par
\REF DLS DLS \Jref\par
\REF FILS FILS \Jref\par
\REF KLS KLS \Jref\par

\font\BF=cmbx10 scaled \magstep 3
\vbox to \vsize{
{\baselineskip=12pt
\line{\hfill 26 June 1994}
\vskip 40pt plus40pt
\centerline{{\BF The Stability of the Peierls Instability}}
\vskip10pt
\centerline{{\BF for Ring-Shaped Molecules}}
\vskip 30pt plus30pt
\centerline{Elliott H. Lieb and Bruno Nachtergaele}
\centerline{Department of Physics}
\centerline{Princeton University}
\centerline{P.O.B. 708, Princeton, NJ 08544}

\vskip 20pt plus40pt

\vfil
\noindent {\bf Abstract}\hfill\break
{
We investigate the conventional tight-binding model of $L$
$\pi$-electrons on a ring-shaped mol\-e\-cule of $L$ atoms with
nearest neighbor hopping.
The hopping amplitudes, $t(w)$, depend on the atomic spacings, $w$,
with an associated distortion energy $V(w)$. A Hubbard type on-site
interaction as well as nearest-neighbor repulsive potentials can
also be included. We prove that when $L=4k+2$ the minimum energy $E$
occurs either for equal spacing or for alternating spacings
(dimerization);
nothing more chaotic can occur.
In particular this statement is true for the Peierls-Hubbard
Hamiltonian which is the case of linear $t(w)$ and quadratic
$V(w)$, i.e.,
$t(w)=t_0-\alpha w$ and $V(w)=k(w-a)^2$, but our results hold
for any choice of couplings or functions $t(w)$ and $V(w)$.
When $L=4k$ we prove that more chaotic
minima {\it can\/} occur, as we show in an explicit example, but the
alternating  state is always asymptotically exact in the limit
$L\to\infty$. Our analysis suggests three interesting conjectures
about how dimerization stabilizes for large systems. We also treat
the spin-Peierls problem and prove that nothing more chaotic than
dimerization occurs for $L=4k+2$  {\it and\/} $L=4k$.
}

\vfil
\noindent {\bf Physics and Astronomy Classification Scheme (1993):}
\hfill\break
\class 31.10.+z  General theory of electronic structure,
                 electronic transitions, and chemical bind\-ing*
\class 31.15.+q  General mathematical and computational
                 developments*
\class 36.20.Hb  Macromolecules and polymer molecules; Configurations
                 (bonds, dimensions)*
\class 63.20.Kr  Phonons and vibrations in crystal lattices;
                 Phonon-electron interactions*
\class 75.10.Lp  General theory and models of magnetic ordering;
                 Band and itinerant models*
\medskip

\hrule width2truein
\smallskip
\noindent
Copyright \copyright\ 1994 by the authors. Faithful reproduction of
this article by any means is permitted for non-commercial purposes.
}
\eject
}
\bgsection 1. INTRODUCTION

To derive the shape (and other properties) of molecules from first
principles has been an actively pursued goal since the early days of
quantum mechanics. Many of the insights into the structure of
molecules like benzene  and its relatives have, however, been obtained
using drastically simplified models. The Schr\"odinger equation for
all the nuclei and electrons in such a molecule involves many dozens or
even hundreds of degrees of freedom and, therefore, simpler models with
a reduced number of degrees of freedom are a necessity. In the case of
benzene the introduction of the H\"uckel model \cite{Hue} played an
important role. This model is standard textbook material for organic
chemistry students (see e.g. \cite{Sal}). London \cite{Lon} used the
H\"uckel model to explain the large diamagnetic anisotropy of aromatic
compounds and certain other materials quite successfully. A similar
approach was used earlier by Jones in his work on bismuth and bismuth
alloys \cite{Jon}.

In this paper we are interested in ring-shaped molecules of the type
$(CH)_{L}$, for {\it even\/} $L$, the so-called {\it annulenes}
(sometimes called cyclic polyenes). The
H\"uckel model for [$L$]-annulene describes the $L$ $\pi-$electrons
(one for each carbon atom) as hopping from one carbon atom to the next
(tight-binding approximation). The carbon atoms are located at the $L$
sites of a ring-shaped geometry, so $L+1\equiv 1$. The Coulomb
interaction between the electrons is ignored in the H\"uckel model but
we will include the Hubbard \cite{Hub} on-site interaction
as in the work of Pariser-Parr \cite{PP} and Pople \cite{Pop},
in our study  (a nearest neighbour repulsion can also be included).
$$
H=-\sum_{j=1}^L\sum_{\sigma=\up,\down} t_j\cd_{j+1,\sigma}\c_{j\sigma}
+ \hc+ U\sum_{j=1}^L (n_{j\up} -\tover12)(n_{j\down}-\tover12)\quad .
\deqno(hubham)$$
Here $\cd_{j+1,\sigma}$ and $\c_{j\sigma}$ are the usual fermion
creation and  annihilation operators for a particle of spin $\sigma$
at site $j$. The number operators are
$n_{j\sigma}=\cd_{j\sigma}\c_{j\sigma}$. The hopping matrix elements
are real, but they can be multiplied by complex phases if a magnetic
field is present. (In \eq(hubham) we write $(n_{j\sigma} -\tover12)$
instead of the customary $n_{j\sigma}$; the difference amounts only to
the addition of trivial terms  in the Hamiltonian, but these are
convenient because they ensure --- via hole-particle symmetry --- that
the expected particle number in the grand canonical ensemble is
$\langle N\rangle=L$, i.e., the ``half-filled band''.) The real
parameter $U$ represents the strength of the on-site interaction of
two particles and the H\"uckel approximation consists in setting $U=0$.
The set of $t_j$'s will be denoted by $\{t_j\}$. The dependence of the
Hamiltonian on the parameters will be made explicit, where needed,
with a notation of the type $H(\{t_j\}), H(\{t_j\},U)$ etc..
For a review of rigorous results on the Hubbard model see \cite{Lie4}.

The hopping matrix element $t_j$ is a resonance integral between
orbitals at the  sites $j$ and $j+1$ and therefore depends on the real
space distance $w_j$ between these sites. A typical choice is
$t_j=t_0-\alpha w_j$ \cite{SSH}. A more realistic choice would be
$t_j=t_0\exp (-\alpha w_j)$ or some other rapidly decaying function of
$w_j$. If one adds to the Hubbard Hamiltonian (even in the H\"uckel
approximation) a term describing the interaction between the carbon
ions (the energy of lattice distortions) one obtains a model that, in
spite of its extreme simplicity, serves very well to explain a number
of phenomena in annulenes and linear $(CH)_x$  (polyacetylene)
\cite{KH}. For a recent review see the references \cite{BCM,BJ}.
Although in more
refined models the lattice distortions should be treated as quantum
mechanical phonons \cite{SSH}, a more common choice,  which we will
follow here, is to describe  them in the Born-Oppenheimer approximation
by adding to the Hamiltonian \eq(hubham) a classical potential of the
form $$
\sum_{j=1}^L V(w_j)
$$
and to minimize, with respect to the $w_j$, the energy functional
$$
\lambda_0(H(\{t(w_j)\})) + \sum_{j=1}^L V(w_j)\quad ,
$$
where $\lambda_0(H)$ denotes the lowest eigenvalue of $H$.
Often $V$ is taken to be quadratic, but in this paper we will {\it
not\/} assume
that $V$ is quadratic or that the dependence of the $t_j$ on the $w_j$
is linear.

The Peierls Instability \cite{Pei}
(discovered also by Fr\"ohlich \cite{Fro} and seemingly
independently by Longuet-Higgins and Salem \cite{LHS})
states that for $L$
sufficiently large and $U$ not too large, the minimum will not be
attained in a translation invariant configuration of $w_j$'s. By a
straightforward computation (exact in the case $U=0$ or perturbative
for small non-zero $U$) one can show that, for any fixed choice of the
function $V$, there is an $L_0$ such that for  all $L\geq L_0$ the
ground state energy of $H$ with
$$
w_j= w_0 +(-1)^j\delta\quad ,
\deqno(dimerw)$$
for small $\delta$, is lower than with the best choice of  $w_j=$
constant.  In the context of $(CH)_x$ molecules this was  discovered
independently by Labhart \cite{Lab} and  Ooshika \cite{Oos1}.

The phenomenon described by \eq(dimerw) is called {\it dimerization\/}
in the physics literature. Unfortunately, this word has quite
different connotations in the chemistry literature, but there does not
seem to be a universally accepted terminology for \eq(dimerw) among
chemists. The phrase {\it bond-alternation\/} \cite{LHS} would be a
more accurate description as far as chemists are concerned. We are
obliged to make a choice here, and we shall use ``dimerization'' ---
in the hope that chemists will substitute ``bond-alternation'' for it
in their minds. In fact we shall go further and {\it declare a
configuration to be dimerized, even if \eq(dimerw) holds with\/}
$\delta=0$, i.e., the $w_j$'s are translation invariant. In other
words, a dimerized configuration is one with period-two  translation
invariance, and this includes period one as a special case.  This
convention, while a bit unusual, conveniently eliminates awkward
locutions.

For the H\"uckel model ($U=0$) and for $L=2\bmod 4$ (i.e.,
$L=6,10,14,\ldots$) the dimerization instability was shown very
explicitly by Longuet-Higgins and Salem who also estimated the degree
of dimerization $\delta$ for realistic values of the parameters
\cite{LHS}. The physical mechanism for the occurence of lattice
distortions is rather simple and quite universal for electron-lattice
systems in one dimension. A lattice distortion of period $1/2k_F$
opens up a gap at the Fermi level, thus lowering the energy of the
occupied levels. This was a basic ingredient in Fr\"ohlich's theory of
superconductivity. The calculations mentioned above show that for
small distortions this  lowering of the energy exceeds the positive
contribution to the total energy of the lattice distortion itself. In
the infinite volume limit the elastic energy per bond is quadratic in
the parameter $\delta$ of \eq(dimerw), while the electronic energy per
bond decreases by an amount proportional to $\delta^2\log \delta$ for
small $\delta$. As far as we know the first author who mentions this
logarithmic behavior is Fr\"ohlich in \cite{Fro}.

A common feature of the works mentioned above is that they always show
that certain instabilities exist. The question of the ``stability of
the instabilities'' is usually not raised, much less resolved.  The
main purpose of this work is to study the stability and instability of
lattice distortions and to obtain rigorous statements about the true
energy minimizing state. Longuet-Higgins and Salem \cite{LHS} raised
the question of the possible occurrence of higher periodicities but
expected them to be unimportant. We prove that in the case $L=2\bmod
4$ (an assumption they made for technical reasons) they were right:
{\it indeed nothing else than periodicity two occurs for $L=2\bmod 4$,
but when $L=0\bmod 4$ other instabilities may, in fact, occur.}

For the H\"uckel model with $t=t_0-\alpha w$ and $V(w)=k(w-a)^2$,
Kennedy and Lieb  \cite{KL} showed that the true minimizing
configuration is {\it always\/} of periodicity two, i.e.,  $w_j$ is of
the form \eq(dimerw). Moreover, the minimizing configuration is unique
(up to translations). It is remarkable that this result holds for {\it
all even\/} $L$, in contrast with what will be proved here for general
functions $V$.

We note parenthetically that the models we study are also used to
describe  electrons in mesoscopic metallic rings \cite{Imr}.
These rings are typically two orders of magnitude larger than the
largest annulenes produced in the laboratory so far, and therefore the
magnetic fields needed to obtain a flux through the ring of the  order
of a flux quantum are much smaller and experimentally accessible. The
main issue is to calculate the persistent currents in such a ring
threaded by a magnetic flux. Depending on the electron number, among
other things, these currents can be paramagnetic or diamagnetic.
Although most authors assume uniform hoppings $t$ for this problem,
there is an intimate connection between these persistent currents and
the stability of dimerized configurations of $t$'s.

In order to proceed to a precise formulation of our results it is
first necessary to confront some possibly confusing questions about
the phases of the $t_j$'s. Let us write
$$
t_j=\vert t_j\vert \exp[i\theta_j]
$$
with $-\pi< \theta_j\leq \pi$. We note, first, that by a simple
unitary gauge transformation of the type $\c_{j\sigma}\to
\exp[i\phi_j]\c_{j\sigma}$ and  $\cd_{j\sigma}\to
\exp[-i\phi_j]\cd_{j\sigma}$, the operators $n_{j\sigma}$ and the
energy levels of $H$ are unchanged, but the $t_j$'s change to  $t_j\to
t_j\exp[i(\phi_j-\phi_{j+1})]$. Consequently, the energy levels depend
{\it only} on the  ``total flux'', $\Phi$, of the $t_j$'s defined by
$$
\Phi= \hbox{argument}(\prod_{j=1}^L t_j)\quad ,
$$
with $-\pi <\Phi \leq \pi$.

The next consideration comes from the physics of our model. The $t_j$'s
are real, unless there is a superimposed magnetic field, in the
absence of which each $\theta_j=0$ or $\pi$, depending on whether
$t_j$ is positive or negative. Both signs can occur. Furthermore, it
is understood that the $t_j$'s remain close to some nonzero value in
the physically realistic situation. Without loss of generality we can
choose this value to be positive. With this in mind, we shall assume
that the elastic energy, as a function of $t$, defined by
$$
f(\vert t\vert)\equiv\inf\{V(w)\mid t(w)=t\}\quad ,
$$
depends only on $\vert t\vert$ and not on the sign of $t$. (We will
impose below in (1.6) some other very mild physical conditions on $f$.)
This is not to say that negative $t$'s will not be permitted. They
will --- as we explain next.

Our procedure will be to study the total energy
$$
\ee_L(\{t_j\})\equiv\lambda_0(H(\{t_j\})) +\sum_{j=1}^L
f(\vert t_j\vert)\quad ,
\deqno(ee)
$$
and to evaluate its minimum
$$
E_L\equiv \min_{\{t_j\}}\ee_L(\{t_j\})\quad ,
\deqno(minee)$$
where the minimum is taken over all choices of $\{\vert t_j\vert\}$
{\it as  well as\/} the phases $\{\theta_j\}$. Thus, we allow negative
$t_j$'s and give them the same elastic energy as $\vert t_j\vert$.
Since the $t_j$'s are not expected to pass through zero in the
physical case, this is no real restriction.
In terms of the $t_j$, dimerization means that the configurations
minimzing $\ee_L(\{t_j\})$ are of the form
$$
t_j=t_0+(-1)^j\delta \quad .
\deqno(dimer)$$

It will turn out that for all $L$ our theorems show that the  energy
minimzing $t_j$'s are real (modulo a gauge transformation). But for
$L=0\bmod 4$ an odd number of them {\it must\/} be negative for an
energy minimizer --- thereby precluding a dimerized state. If, on the
other hand, we wish to attribute the  occurrence of an odd number of
negative $t_j$'s to the presence of an external magnetic field then we
can say that a dimerized state minimizes the energy in the presence of
a field of flux $\pi$. Given that no such field is really present, we
will have to conclude that the energy mimimum will not be a dimerized
state when $L=0\bmod 4$. In this case we can redefine our problem by
restricting the minimum in \eq(minee) to positive $t_j$'s. Do we get a
dimerized state then? We do not know the general answer but we have
some conjectures about this question. If $f(t)=k(t-a)^2$, as in
\cite{KL}, we know that a dimerized state is, indeed, the minimum.

The physical conditions we impose on $f$ are
$$\eqalign{
&\hbox{i) $f$ is continuous. \qquad\qquad\qquad\hfill}\cr
%&\phantom{a} \cr
&\hbox{ii) $f(t)\geq Ct$, for large $t$ and for some
constant $C>4$. \qquad\qquad\qquad\hfill }\cr
}\deqno(conditions)$$
The continuity of $f$ is not really necessary for the validity of our
results. In fact in the Appendix (\Lem13/convexf/) we show that $f$
can always be
replaced with a convex (and hence continuous) function  without
changing the minimizing configurations. Property ii) is necessary to
have a stable minimum at all, e.g.,  to prevent the molecule from
collapsing to a point.

We minimize the ground state energy $\ee_L$ not only with respect to
the parameters $t_j$ but {\it also\/} with respect to the number of
electrons.  In Section 2 we prove that if $U\neq 0$ and all $t_j$
are real, the ground states of \eq(hubham) will in fact necessarily
have $N=L$. In this case all results below are equally valid for
minimization of the energy with the particle number fixed at
half-filling. The condition $U\neq 0$ is necessary in the case of
rings of size $L=0\bmod 4$, for with $U=0$ the model with translation
invariant $t$'s has ground states with particle number ranging from
$L=-2$ to $L+2$, due to zero eigenvalues of the one-particle
Hamiltonian.

\medskip
\noindent
{\bf Relation with the Flux-Phase Problem.}\nl
The minimization with respect to the flux $\Phi$ of $\ee_L(\{t_j\})$,
for fixed $\vert t_j\vert$, is a generalization of the so-called flux-phase
problem \cite{Wie,Lie3} for rings. In the case $U=0$ (as in the
original  formulation of this problem),  but arbitrary, fixed,
positive $\{t_j\}$, this problem was solved by Lieb and Loss
\cite{LS}. The arguments of this paper provide a way of solving the
problem for rings and all values of $U$ (\Cor8/fluxphase/).  For rings
of length $L=2\bmod 4$ the optimal ($\equiv$ energy minimizing) flux
is 0. When $L=0\bmod 4$, the optimal flux is $\phi=\pi$. This
can be extended to  higher dimensions \cite{Lie2}
when some geometric
periodicity is present. In particular, \cite{Lie2} proves the
conjecture that  the optimal flux for the Hubbard model on the
two-dimenisonal square lattice is $\pi$ through each plaquette.

\medskip
\noindent
{\bf Main Results.\/}\nl
The following theorem, which we prove in Section 3,  says that, when
$L=2\bmod 4$ and the flux is zero and when  $L=0\bmod 4$ and the flux
is $\pi$, the Peierls instability is itself stable in the sense that
no other lattice distortions than period two can occur.  The theorem
does {\it not\/} state that distortion always happens. Depending on
the parameters in the Hamiltonian, the function $f$ in particular, the
energy minimzing configuration may or may not be translation
invariant. In real molecules both situations occur. Benzene, e.g., has
all $C-C$ bond lengths equal (within the precision of today's
measurements and in agreement with the analysis of Labhart \cite{Lab}
and Ooshika \cite{Oos2}), while  for large rings or long chains the
Peierls instability will necessarily lead to  two different bond
lengths.

\iproclaim/global/ Theorem (Minimizing configurations).
\item{i)} If $L=2 \bmod 4$, the minimum of $\ee_L$ is attained in a
dimerized configuration of the form \eq(dimer) in which the $t_j$'s
are all positive (or all negative).
\item{ii)} If $L=0\bmod 4$, the minimum of $\ee_L$ is attained
configuration in which the $\vert t_j\vert$'s satisfy \eq(dimer) but
the flux $\Phi$ equals $\pi$.
\eproclaim

That the cases $L=2\bmod 4$ and $L=0\bmod 4$ behave differently with
respect to the flux was noted a long time ago (see e.g. the ``H\"uckel
rule'' in \cite{Sal}.)  The $U=0$ one-electron problem has zero-energy
states when $L=0\bmod 4$ but not when $L=2\bmod 4$. In the first case
this leads  to diamagnetic, in the second case to paramagnetic
response to magnetic fields, as was noted in \cite{Lon} and as is
observed experimentally in the annulenes  \cite{Gar}.

\iproclaim/uniquedim/ Theorem (Uniqueness).
If there is only one (up to translation and gauge transformations)
dimerized configuration satisfying the conclusion of \Thm/global/,
then there are no other energy minimizing configurations.
\eproclaim

We define three additional minimization problems for the $L=0\bmod 4$
case.

\item{$\bullet$} {\it Zero-flux problem:\/}  We restrict the
minimization in \eq(minee) to those $\{t_j\}$ with zero flux, i.e.,
$\Phi=0$.
\item{$\bullet$} {\it Period-two problem:\/} In addition to $\Phi=0$
we make the further restriction that the $t_j$'s must be dimerized.
\item{$\bullet$} {\it The infinite-chain problem:\/} Let $d(x,y)$
denote the dimerized configuration with $t_{2j+1}=x$ and
$t_{2j}=y$, and minimize the energy per site defined by
$$
e(x,y)=\lim_{L\to\infty} {1\over L} \ee_L(d(x,y))\quad .
\deqno(density)$$

The minimum energy in the first case is denoted by $E^0_L$, and the
$t_j$'s by $t^0_j$. The second energy is denoted by
$E^0_{L,\hbox{\eightrm dimer}}$ and the $t_j$'s by
$t^0_{j,\hbox{\eightrm dimer}}$. Clearly
$$
E^0_{L,\hbox{\eightrm dimer}} \geq E^0_L
$$
and one of our conjectures below concerns the case of equality. The
following, however, shows that the three problems become asymptotically
the same (to within two powers of $L^{-1}$, instead of merely one) as
$L\to\infty$.

\iproclaim/asymptotics/Theorem (Asymptotic dimerization).
\item{i)}
Assume that there is a unique energy minimizing
dimerized configuration $d(x,y)$, $x\geq y$, for the
infinite-chain problem. Then any sequence of minimizing
configurations $\{t^0_j\}$ of the zero-flux problem converges to a
dimerized configuration as $L\to\infty$. Namely,
$$
\lim_{L\to\infty} \max_j \vert t^0_j - t^0_{j,\hbox{\eightrm dimer}}
\vert =
\lim_{L\to\infty} \max_j \vert t^0_j - d(x,y)_j
\vert = 0 \quad ,
$$
where we have adopted the convention that $t_1$ is the largest of the
$t_j$'s for all configurations.
\item{ii)}
For any even $L$, the energies satisfy
$$
0\leq E_{L,\hbox{\eightrm dimer}} - E^0_L
\leq \hbox{constant}\times L^{-1}\quad .
$$
\eproclaim

In section 5 we discuss the situation of finite rings of length $L=
0\bmod 4$. Instabilities other than period-two can exist for
arbitrarily large rings if the function $f$ is chosen appropriately.
An optimally dimerized ring of size $L=0\bmod 4$ can have other
instabilities and it seems that the strongest of those actually has
wavelength $=L$, and is not, as one could have expected,  of period 4
or some other higher periodicity. We illustrate this with an explicit
example of a ring of 8 sites for which the minimum is not period 2. A
perturbative analysis indicates that the conditions on $f$ for a
further instability to occur become more stringent as the system
becomes larger. In particular the dimerization itself has to be
smaller as $L$ becomes bigger. This leads us to the following two
conjectures:

\smallskip
\noindent
{\bf Conjecture 1.} For every $L=0\bmod 4$ there exists a function $f$
(depending on $L$), satisfying conditions i) and ii)
in \eq(conditions),
such that the energy minimizing configurations for the zero-flux
problem are {\it not\/} period two.

\smallskip
\noindent
{\bf Conjecture 2.} For any fixed $f$ satisfying \eq(conditions),
and such
that there is a unique minimum for the infinite-chain problem, there is
a critical size $L_c$ such that for all even $L\geq L_c$ all zero-flux
minima are  period two.

\medskip
\noindent
{\bf The Spin-Peierls Problem.}\nl
Another closely related question is the spin-Peierls problem. For $U$
large, and half-filling, the Hubbard model reduces to the Heisenberg
antiferromagnetic spin chain in second order perturbation theory
\cite{And}. The spin chain also has an instability when coupled to
lattice distortions, as was pointed out by Chesnut \cite{Chs} and
by Beni and Pincus \cite{Pin,BP}. More explicitly one looks for
the  configurations of coupling constants $J_{ij} $ that minimize the
smallest eigenvalue of the Hamiltonian:
$$
H_\Lambda=\sum_{<i,j>} J_{ij} \es_{i}\cdot\es_{j} +
\sum_{<i,j>} f(J_{ij})\quad ,
\deqno(heisham)$$
where the sum is over nearest neighbour pairs in a finite subset
$\Lambda$ of  the hypercubic lattice, and $f(J_{ij})$ is the elastic
energy as before. In one dimension Cross and Fisher \cite{CF} computed
the exponent  governing the ground state energy density $e(\delta)$ of
the system with  alternating couplings:
$$
J_j =1+(-1)^j \delta\quad ,
$$
with $J_j\equiv J_{j,j+1}$.
They found that, up to logarithmic corrections,
$$
\vert e(\delta)-e(0)\vert\sim -\vert\delta\vert^{4/3}\quad .
$$
Numerical confirmation of this exponent and evidence for a
logarithmic correction was reported in \cite{SKM}.
The important point here is that  $4/3 <2$, which implies the
instability  of the translation invariant configurations under period
2 perturbations, provided $f$ has a finite second derivative. It was
recently proved that this exponent is identical to a critical exponent
of the two dimensional 4-state Potts model \cite{AN}.

The question addressed in this paper is again to determine the nature
of the minimizing configurations. It turns out that the true minimum
is dimerized for any (even) system size  and in any dimension. {\it
The $2 \bmod 4$ versus $0 \bmod 4$ dichotomy does not arise for the
Heisenberg model!}

\iproclaim/spinpeierls/ Theorem (Dimerization for spin-Peierls).
Let $\Lambda\subset \Ir^d$ be a rectangular box of even size in all
coordinate directions  and with periodic boundary conditions.  Then
there is an energy minimizing configuration of $J_{ij}$'s  which is of
periodicity 2 in all coordinate directions. \eproclaim

This theorem prompts one more conjecture.

\smallskip
\noindent
{\bf Conjecture 3.} Suppose that a minimizer for the zero-flux problem
for some $f$ and $L=0\bmod 4$ and some $U_0\geq 0$ is a dimerized
state. Then for every $U > U_0$ there is a dimerized minimizer for the
zero-flux problem with the same $f$ and $L$. In particular, the truth
of this conjecture, when combined with the $U=0$ result in \cite{KL},
would imply that, with a quadratic $f$, dimerization always occurs,
even in the $L=0\bmod 4$ case.

\medskip
\noindent
{\bf Extensions of the results presented in this paper.}

1) It has been argued that the Hubbard on-site repulsion does not
describe the Coulomb interaction between the electrons accurately
enough (for a discussion see e.g. the reviews on this topic in
\cite{BC} and references therein). While the Hubbard term alone, when
not too large, seems to have the effect of enhancing the dimerization
\cite{DM},
taking into account some nearest neighbour interaction terms Kivelson,
Su, Schriefer, and Heeger \cite{KSSH} found that the  Coulomb
repulsion in fact suppresses the dimerization in polyacetylene.
Without entering into this discussion here we would like to point out
that without substantial modification  one can prove the results of
\Thm/global/ and \Thm/uniquedim/ for a Hamiltonian that includes a
repulsive nearest-neighbour interaction  of the form
$$
\sum_j W(t_j) (n_j-1)(n_{j+1}-1)
$$
The $-1$'s are inserted to keep the chemical potential tuned at
half-filling. We require $W(t)\geq 0$, but no assumption on the
dependence of $W$ on $t$ is needed. Theorem 1 holds unchanged, but we
can extend the proof  that the grand canonical ground states are
necessarily half-filled only under the additional condition that
$W(t_{j-1}) + W(t_j) < U_j$, for all $t$'s in the relevant range.
Nearest neighbour exchange terms and certain longer range interactions
can also be treated.

2) Instead of a homogeneous on-site repulsion it is sometimes more
realistic to have different $U$ for different (classes of) sites,
e.g. in a periodic manner. Theorem 1 extends to such situations in the
following  sense: the minimum energy will be attained in a
configuration of $t_j$  which is invariant under all reflections
through planes that intersect two oppossite bonds on the ring (as
indicated by a dashed line in Figure 1) and that leave the
configuration of the $U_i$ invariant. As an example consider
[18]-annulene. Of the 18 hydrogens in this molecule 6 reside inside
the ring and 12 outside, according to the pattern: 1 inside, 2 outside,
1 inside, etc. A possible way of taking the effect of this periodic
configuration of H-atoms into account, is to add a periodic
one-particle potential to the Hamiltonian. In the present example this
potential would take two different values  and depend on the site
according to the pattern: $(v_i)=(v_1,v_1, v_2, v_1, v_1, \ldots)$.
Our general result then implies an energy minimizing configuration of
$C-C$ bonds of the form $ABCACBABCACB\ \cdots$. The observed bond
lengths  in [18]-annulene indeed satisfy this pattern (they are
$A=1.419$\AA, $B=1.382$\AA,$ C=B$ \cite{Gar}.)

3) Hubbard models with spin-dependent hoppings, i.e., $t_j= \vert
t_j\vert m_j, m_j\in$ SU(2), have been considered in the literature in
order to study the effects of spin-orbit coupling \cite{MGEW,EWGMO,FK}.
By virtue of the SU(2) gauge symmetry of the Hamiltonian one can
diagonalize the spin dependence of the hoppings up to an SU(2) flux,
which is defined up to an SU(2) transformation by the matrix $M=m_1
m_2 m_3 \cdots m_L$. The methods of this paper can be used to prove
the existence of a minimizer having dimerization and $M=\idty$ when
$L=2\bmod 4$ and when  $L=0\bmod 4$ a minimizer has $M=-\idty$ and
dimerization. This is completely analogous to the U(1) case treated in
\Thm/global/.

4) Mattis and Langer \cite{ML} introduced a simple model to study  the
interaction between the electrons and a phonon instability of the
Peierls type as a function of the temperature. Our methods can also be
used to prove that in this model the electron state has periodicity
two and that its correlations indicate dimerization. For a rigorous
study of the Kohn anomaly, which accompanies the  phase transition in
the Mattis-Langer model, see \cite{PVZ}.

Generalizations that are not considered in this work include the
interaction between the electrons and lattice in excited states, the
effects of doping, and higher dimensional models. Low-lying
electronic excited states can induce further (e.g.
soliton-like) distortions of the lattice \cite{SSH,SG,HM}. In doped
polyacetylene an interesting semiconductor-metal transition occurs
\cite{MR}. Away from half-filling, instabilities other than period-two
will naturally develop (see e.g. \cite{UMT} and references therein).
In two dimensions, e.g., on a square lattice, $2k_F$ instabilities may
develop either in the coordinate  directions or in the diagonal
directions \cite{Maz,TH}. This may be related to the non-period-two
instabilities that can occur in rings whose size is a multiple of 4,
e.g., in the elementary plaquettes of the square lattice. Another new
feature of higher dimensions is that the breaking of translation
invariance can occur at non-zero temperature. This has been rigorously
shown to occur in the Falicov-Kimball model \cite{KL2,BS} and in the
Holstein model \cite{LM}.

\bgsection 2. THE GRAND CANONICAL HUBBARD MODEL AT HALF-FILLING

In the definition of the energy functional \eq(ee) we used the  lowest
eigenvalue of the grand canonical Hubbard Hamiltonian at half-filling.
Our goal in this section is to prove that the corresponding ground
state necessarily has $N=L$ if we assume the $t_j$'s are real (as will
later be shown to be the case). It is then obvious that minimization
of the energy with respect to the parameters $t_j$ {\it and\/} the
number of electrons, produces exactly the same minimizing
configurations  as the restricted mimimization with the particle
number fixed at $N=L$. The arguments in this section are not restricted
to one dimension. We believe they are interesting in their own right
and can be of interest in a more general context. Therefore, we
temporarily consider a more general setup on a general finite lattice
$\Lambda$, the number of whose lattice sites is denoted by
$\vert\Lambda\vert$.

Consider the grand canonical Hubbard Hamiltonian \eq(hubham) for spin
1/2 fermions on $\Lambda$, and with arbitrary
chemical potential $\mu$:
$$
H=-\sum_{i,j\in\Lambda}\sum_{\sigma=\uparrow,\downarrow}
t_{ij}\cd_{j\sigma}\c_{i\sigma}
+\sum_{j\in \Lambda} U_j(n_{j\uparrow}-\mu)(n_{j\downarrow}-\mu)
$$
The $t_{ij}$ are asssumed (in this section only) to be real and
symmetric, i.e
$t_{ij}=t_{ji}$.
$\Lambda$ is assumed to be connected, i.e., given $i$ and $j$ in
$\Lambda$ we can find a sequence $i=i_1,i_2,\ldots,i_m=j$ such that
$t_{i_1,i_2}t_{i_2,i_3}\cdots t_{i_{m-1},i_m}\neq 0$.
The $U_j$ and $\mu$ are also real parameters. As
before the dependence of $H$ on its parameters will be made explicit
with a notation of the type $H_\mu$, $H(\{t_{ij}\})$,
$H_\mu(\{t_{ij}\}, \{U_j\})$ etc. When $H$ appears without subscript
the chemical potential is $\mu=1/2$, corresponding to half-filling.

The spin-up and spin-down particle number operators are defined by
$$
N_\sigma=\sum_{j\in\Lambda}n_{j\sigma}
$$
where $\sigma=\uparrow,\downarrow$ and $n_{j\sigma}=
\cd_{j\sigma}\c_{j\sigma}$. Note that $N_\uparrow$ and $N_\downarrow$
each commute with $H$, and hence their eigenvalues are good quantum
numbers. $H$ also commutes with the total-spin operators (which
generate the global spin rotations) given by
$$
S^+=\sum_j \cd_{j\up}\c_{j\down},\quad S^-=\sum_j \cd_{j\down}
\c_{j\up},\quad
S^3={1\over 2}(N_\up-N_\down)
$$
Therefore, by applying $S^+$ or $S^-$ an appropriate number of times,
one can transform any ground state of $H$ into a ground state with
$$
N_\up-N_\down=\cases{0& if $N$ is even\cr
               \pm 1 & if $N$ is odd\cr}
\deqno(diffN)$$

We say that $\Lambda$ and $t$ are bipartite if the sites of
$\Lambda$ can be written as the
disjoint union of two subsets; i.e., $\Lambda=\Lambda_A
\cup\Lambda_B$, in such a way that the matrix $t$
of hopping matrix
elements respects this structure, namely
$t_{ij}=0$
if $i\in \Lambda_A$ and $j\in \Lambda_A$
or if $i\in \Lambda_B$ and $j\in \Lambda_B$. In this case
the Hamiltonians $H_\half(\{t_{ij}\},\{U_j\})$ and
$H_\half(\{t_{ij}\},\{-U_j\})$ are unitarily equivalent, i.e at
half-filling ($\mu=\half$) a global sign change of the potential does
not affect the spectrum. A unitary transformation  implementing this
equivalence is the particle-hole transformation mapping
$$\eqalign{
\cd_{j\uparrow}\longrightarrow \epsilon(j) \c_{j\uparrow},& \quad
\c_{j\uparrow}\longrightarrow \epsilon(j) \cd_{j\uparrow}\cr
\cd_{j\downarrow}\longrightarrow \cd_{j\downarrow},& \quad
\c_{j\downarrow}\longrightarrow \c_{j\downarrow}\cr
}$$
where $\epsilon(j)=1$ for $j\in\Lambda_A$ and
$\epsilon(j)=-1$ for $j\in\Lambda_B$.
By the same transformation the spin operators are mapped
into the so-called {\it pseudospin\/} operators defined by
$$
\tilde S^+=\sum_j \epsilon(j)\c_{j\up}\c_{j\down},\quad
\tilde S^-=\sum_j \epsilon(j)\cd_{j\down}\cd_{j\up},\quad
\tilde S^3={1\over 2}(\vert\Lambda\vert-N_\up-N_\down)
$$
These operators therefore also commute with the Hamiltonian, and they
also commute with the spin operators and generate an additional SU(2)
symmetry. Under the particle-hole transformation $N_\up$ and $N_\down$
are transformed into $\vert\Lambda\vert- N_\up$ and $N_\down$
respectively. It is then obvious that by applying $\tilde S^\pm$ to
any ground state we can obtain a ground state with
$$
N_\up+N_\down=\cases{\vert\Lambda\vert
& if $N-\vert\Lambda\vert$ is even\cr
              \vert\Lambda\vert \pm 1
& if $N-\vert\Lambda\vert$ is odd\cr}
\deqno(sumN)$$

As the spin and pseudospin operators commute \eq(diffN) and  \eq(sumN)
can be realized simultaneously. For an even bipartite lattice and
$\mu=1/2$ this implies that there is a ground state of $H$ with
$N_\up=N_\down=\vert\Lambda\vert/2$ or  $N_\up=N_\down \pm
1=\vert\Lambda \vert/2$.

{}From the next lemma it will follow that if in addition all $U_j$ are
nonvanishing and of the same sign, {\it all\/} ground states have even
particle number and satisfy  $N_\up=N_\down=\vert\Lambda\vert/2$. Note
that the Lemma itself does not require $\Lambda$ to be bipartite.

\iproclaim/spin0/ Lemma.
If all $t_{ij}$ are real and $U_j\leq 0$ for all $j$, then the ground
state space of $H(\{t_{ij}\},\{U_j\})$ contains a state with
$N_\uparrow=N_\downarrow$. If $U_j< 0$ for all $j$, all groundstates
satisfy $N_\uparrow=N_\downarrow$. In particular the total number of
particles in the ground state is {\it even} and the total spin is
{\it zero}.
\eproclaim
\proof:
We can introduce new operators $\hat\c_{i,\sigma}$ defined by
$$
\hat\c_{j,\up}=\c_{j,\up},\quad \hat\c_{j,\down}=(-1)^{N_\up}
c_{j,\down}
$$
$H$ has then the same form in terms of the $\tilde\c$'s as in terms of
the $\c$'s, but now the $\hat c^\#_{j,\down}$ commute with the $\hat
c^\#_{j,\up}$.

Our Hamiltonian is now an operator on $\H_\up\otimes\H_\down$, and can
be written in the form
$$
H=T\otimes\idty +\idty\otimes T + \sum_{i\in\Lambda} U_j
(n_j-\mu)\otimes (n_j-\mu)
$$
where $T=-\sum_{i,j\in\Lambda} t_{ij} \cd_j\c_i$. (Because spin $\up$
and spin $\down$ operators act identically on different tensor
factors in the Hilbert space, we can omit the spin index.) $T$ and the
$n_j$ are real in the canonical basis of localized particles.
Therefore we can apply \Lem14/DLS/ of the appendix and the discussion
thereafter to conclude that $H$ has a ground state with
$N_\up=N_\down$.

If $U_j<0$ for all $i$, there cannot be a ground state with $N_\up\neq
N_\down$. Indeed the last statement of \Lem14/DLS/ implies that  for
any ground state $\Omega$
$$
U_j\langle \Omega \mid (n_{j,\up} -\mu)\Omega\rangle  = U_j\langle
\Omega \mid (n_{j,\down} -\mu)\Omega\rangle
$$
If $U_j<0$ this implies $N_\up=N_\down$.
\QED

\iproclaim/uniqueness/ Lemma.
If all $t_{ij}$ are real and bipartite, $\vert\Lambda_A\vert
=\vert\Lambda_B\vert$,  and either all $U_j>0$ or all $U_j<0$.  Then
the ground state of $H$ with $\mu=1/2$ is  unique and has
$\Nup=\Ndown=\vert\Lambda\vert/2$.
\eproclaim
\proof:
It is sufficient to consider the case of all $U_j>0$. The case
$U_j<0$ then follows because the particle-hole transformation changes
the sign of the $U_j$, and the properties $\Nup=\vert\Lambda\vert/2$
and  $\Ndown=\vert\Lambda\vert/2$ are unchanged under the
transformation. Let us therefore consider the case $U_j>0$, for all
$j$.

We first apply \Lem/spin0/ to the particle-hole transformed
Hamiltonian which has all $U_j<0$. This tells us that
$N_\down=\vert\Lambda\vert-N_\up$ in all ground states. In particular
the total number of particles is  $\vert\Lambda\vert$ which, by
assumption, is even.  Then, by Theorem 2 of \cite{Lie1} (the
grand-canonical version at half-filling does not need the homogeneity
of the potential) and the assumptions on the lattice,
$\Nup=\Ndown=\vert\Lambda\vert/2$. [Note: Theorem 2 of \cite{Lie1}
requires $U_j=$ constant. That is the  case when the interaction is
$n_{j\up}n_{j\down}$ as in \cite{Lie1}. If we use the formulation of
\eq(hubham), however, with
 $(n_{j\up}-1/2)(n_{j\down}-1/2)$, the same proof works without the
need of constant $U_j$.]
\QED

For translation invariant $t_{ij}$ on regular lattices the case
$U_j\equiv 0$ is of course exactly solvable. It is rather
straightforward in such cases to determine the ground state degeneracy
``by hand''. The ground state is unique if and only if the eigenvalues
of the matrix $(t_{ij})$ are all non-zero.

It is straightforward to extend \Lem/spin0/ to Hamiltonians that
include interactions of the form
$$
F(\{n_{j,\up}\}) + F(\{n_{j,\down}\})
-\sum_\alpha G_\alpha(\{n_{j,\up}\}) G_\alpha(\{n_{j,\down}\})
$$
for arbitrary real functions $F$ and $G$ of the local particle numbers.
One can use this to prove a slightly weaker form of \Lem/uniqueness/
for a class of models with an additional term of the form
$$
-\sum_{ij} W_{ij}(n_i-1)(n_j-1)
$$
added to the Hamiltonian $H$. For these models our methods also prove
that

i) if $W_{ij} \geq 0$ and $(U_j-\sum_{i} W_{ij})\geq 0$, for all $j$,
then there is a ground state with $N_\up=N_\down=\vert\Lambda\vert/2$.

ii) if $W_{ij} \geq 0$ and $(U_j-\sum_{i} W_{ij})> 0$, all ground
states have $N_\up=N_\down=\vert\Lambda\vert/2$.

\bgsection 3. GENERAL RESULTS BASED ON REFLECTION
POSITIVITY;\hfill\break
PROOFS OF THEOREMS 1 AND 4

We now return to the half-filled Hubbard model ($N=\vert\Lambda\vert$)
on a ring of even length $\vert\Lambda\vert=L$.  It will be convenient
to express the dependence of the energy functional on $\{t_j\}$ in a
more explicit way. We write
$$
\ee_L(\tl,\tm,\tr)=
\lambda_0(H(\{t_j\})) +\sum_{j=1}^L f(\vert t_j \vert)\quad ,
$$
where $\tl,\tr$, and $\tm$ denote a partition of the {\it complex\/}
$t_j$'s (with $t_j^*$ denoting the complex conjugate of $t_j$)
into three groups as follows (also see Figure 1):
$$
\tr=(t_1,\ldots,t_{L/2-1}),\quad
\tm=(t_L, t_{L/2}),\quad
\tl=(t_{L-1},\ldots,t_{L/2+1})\quad .
$$
The reflection of a configuration $(t^{(l)},t^{(m)},t^{(r)})$ through
the plane intersecting the bonds $\{L,1\}$ and $\{L/2,L/2 +1\}$ is the
configuration $(t^{(r)},t^{(m)},t^{(l)})$, i.e., with
$t^{(l)}\leftrightarrow t^{(r)}$.

Throughout this section we assume a constant potential $U_j=U$. The
following lemma can be formulated to include non-constant potentials
but this is not needed for our purposes.

\iproclaim/RPineq/ Lemma.
Assume a constant potential $U_j=U$. If $L=2\bmod 4$, assume that
$t_{L/2}$ and $t_L$ are real and nonnegative. If $L=0\bmod 4$, assume
that $t_{L/2}\geq 0$ and $t_L\leq 0$. The other $t_j$ are allowed to
be arbitrary complex numbers. Then the energy functional satisfies
$$
\ee_L(\tl,\tm,\tr)\geq {1\over 2} \left(\ee_L(\tl,\tm,\btl)+
\ee_L(\btr,\tm,\tr)\right)\quad .
\deqno(RPineq)$$
\eproclaim
\proof:
As in the proof of \Lem/spin0/ we consider the Hamiltonian \eq(hubham)
as acting on $\H_\uparrow\otimes\H_\downarrow$, which amounts to
considering operators referring to opposite spins as commuting. For
$\sigma=\uparrow,\downarrow$ define a set of ``spin'' operators using
a Jordan-Wigner transformation as follows:
$$\eqalign{
S^3_{j,\sigma}&=\cd_{j,\sigma}\c_{j,\sigma} -\tover12\quad ,
\quad  \es_{j,\sigma}=2S^3_{j,\sigma}\quad ,\cr
S^+_{j,\sigma}&= \es_{1,\sigma}\cdots\es_{j-1,\sigma}
\cd_{j,\sigma}\quad ,
S^-_{j,\sigma}= \es_{1,\sigma}\cdots\es_{j-1,\sigma}
\c_{j,\sigma}\quad .\cr
}$$
With these definition and the anticommutation relations of the
$\c_{i,\sigma}$ and $\cd_{i,\sigma}$ it is straigthforward to check
that $S^3_{i,\sigma},S^+_{i,\sigma}$, and $S^-_{i,\sigma}$ satisfy the
standard SU(2) commutation relations and that they commute for
different indices $i,\sigma$. In terms of these operators,
and using the commutation relations, the Hamitonian can be written as
$$\eqalign{
H&=\sum_{j=1}^{L-1}\sum_{\sigma=\uparrow,\downarrow}
t_j S^+_{j+1,\sigma}S^-_{j,\sigma}+ \hc
+ U\sum_{j=1}^L S^3_{j,\uparrow}S^3_{j,\downarrow}\cr
&\quad t_L\sum_{\sigma=\uparrow,\downarrow}
S^+_{1,\sigma}(\es_{1,\sigma}\cdots\es_{L,\sigma})S^-_{L,\sigma}
+ \hc \quad .\cr
}$$
By performing a rotation by $\pi$ about the 2-axis at every other
site and for both $\sigma=\uparrow$ and $\sigma=\downarrow$ we find
that this spin Hamiltonian is unitarily equivalent to:
$$\eqalign{
\tilde H&=\sum_{j=1}^{L-1}\sum_{\sigma=\uparrow,\downarrow}
-t_j S^+_{j+1,\sigma}S^+_{j,\sigma}+ \hc
+ \sum_{j=1}^L U_j  S^3_{j,\uparrow}S^3_{j,\downarrow}\cr
&\quad + (-1)^{L/2} t_L\sum_{\sigma=\uparrow,\downarrow}
S^+_{1,\sigma}(\es_{1,\sigma}\cdots\es_{L/2,\sigma})
S^+_{L,\sigma}(\es_{L,\sigma}\cdots\es_{L/2+1,\sigma}) + \hc \quad .\cr
}$$
Under the conditions stated in the proposition $t_{L/2}\geq 0$ and
$(-1)^{L/2} t_L\leq 0$. We can now apply \Lem14/DLS/ of the
appendix to $\tilde H$ with
$$\eqalign{
A&=-\sum_{j=1}^{L/2-1}\sum_{\sigma=\uparrow,\downarrow}
t_j S^+_{j+1,\sigma}S^+_{j,\sigma}+ \hc
+ U \sum_{j=1}^{L/2}S^3_{j,\uparrow}S^3_{j,\downarrow}\quad ,\cr
B&=-\sum_{j=L/2+1}^{L-1}\sum_{\sigma=\uparrow,\downarrow}
t_j S^+_{j+1,\sigma}S^+_{j,\sigma}
+ \hc + U\sum_{j=L/2 +1}^L S^3_{j\up}S^3_{j\down}\quad ,\cr
C_1\otimes C_1&= t_{L/2} S^+_{L/2 +1,\up}S^+_{L/2,\up}\quad ,\cr
C_2\otimes C_2&= \vert t_L\vert S^+_{1,\up}(\es_{1,\up}
\cdots\es_{L/2,\up})
S^+_{L,\up}(\es_{L,\up}\cdots\es_{L/2+1,\up})\quad ,\cr
}$$
and six more terms $C_i\otimes C_i$, $i=3,\ldots,8$, which are equal
to $C_1\otimes C_1$ and $C_2\otimes C_2$ with $\up$ replaced by
$\down$ and the hermitian conjugates of these operators. This gives us
the inequality of the lemma for the lowest eigenvalue of $H$. The
elastic energy terms in the functional $\ee_L$ add up to the same
contribution on both sides of the inequality and therefore can be
included trivially.
\QED

\noindent
{\bf Proof of Theorem 1 :}\nl
First we prove that the minimum of $\ee_L$ is attained in a
configuration of $t_j$ with total flux $\Phi$ when $L=2\bmod 4$,
and with $\Phi=\pi$ when $L=0\bmod 4$. Let $\{t_j\}$ be minimizing.
Due to the invariance of the ground state energy of $H$ under gauge
transformations, the energy functional $\ee_L$ defined in \eq(ee)
depends only on $\{\vert t_j\vert\}$ and $\Phi$. Hence, for any set of
$t_j$'s there are $t_j^\prime$ such that $\ee_L(\{t_j\})=
\ee_L(\{t_j^\prime\})$ and  $t_{L/2}^\prime$ and $t_L^\prime$ are
real and have the correct sign for application of \Lem/RPineq/. The
two configurations in the right side of \eq(RPineq) have flux $\Phi=
\hbox{argument}(t_{L/2}^\prime t_L^\prime)$ which, by assumption, takes
the values stated in the theorem. Because $\{t_j\}$ is minimizing,
\eq(RPineq) must be an equality. Therefore the two configurations
on the right side must be minimizers of $\ee_L$, but then both have
the correct flux, $0$ or $\pi$.

Next we show that in both cases the
minimum of $\ee_L$ is attained  in a configuration with $\{\vert
t_j\vert\}$ dimerized. For any configuration $(\tl,\tm,\tr)$ of real
$t_j$'s and $\tm$ of the right signs, \eq(RPineq) implies that either
$(\tl,\tm,\tl)$ or $(\tr,\tm,\tr)$ has at least as low an energy and
has at least as many pairs of identical $\vert t_j \vert$'s. In
particular this shows that $\ee_L$ is minimized in a configuration
with $\vert t_j \vert =\vert t_{j+2}\vert $ for $j=L/2-1$ and $j=L-1$.
Because $\ee_L$ is translation invariant the argument above can be
repeated to show that there is a minimizing configuration with $\vert
t_j \vert =\vert t_{j+2} \vert$ for all $j$, which is the desired
result.
\QEDnogroup

\Lem/RPineq/ also permits us to solve the flux phase problem for even
rings with $U\neq 0$.

\iproclaim/fluxphase/Corollary.
Let $U_j\equiv U$ and let $\{t_j\}$ be a fixed  configuration of
nonnegative $t_j$'s. Then the minimum of $\ee_L$,  varying over the
total flux $\Phi$ alone, is attained for $\Phi=0$ if $L=2\bmod 4$ and
for $\Phi=\pi$ if $L=0\bmod 4$.
\eproclaim

\noindent
{\bf Proof of Theorem 4 :}\nl
The Heisenberg antiferromagnet on an arbitrary bipartite lattice is
reflection positive \cite{DLS}. This implies, via \Lem14/DLS/, the
following inequality, analogous to \Lem/RPineq/:
$$\eqalign{
&\lambda_0(H_{\hbox{\eightrm Heis.}}(\{J_{ij}^{(l)},
J_{ij}^{(m)},J_{ij}^{(r)}\})
\cr
&\quad \geq {1\over 2}\left(
\lambda_0(H_{\hbox{\eightrm Heis.}}(\{J_{ij}^{(l)},
J_{ij}^{(m)},J_{ij}^{(l)}\})
+\lambda_0(H_{\hbox{\eightrm Heis.}}(\{J_{ij}^{(r)},
J_{ij}^{(m)},J_{ij}^{(r)}\})
\right)\quad,\cr
}$$
where $\{J_{ij}^{(l)},J_{ij}^{(m)},J_{ij}^{(r)}\}$ denotes a partition
of the coupling constants into three groups: the $J_{ij}^{(l)}$ are
the couplings on the bonds to the left of any reflection plane of the
lattice, the $J_{ij}^{(m)}$ are the couplings on the bonds intersected
by the reflection plane, and the $J_{ij}^{(r)}$  are the coupling to
the right of the plane. Using this inequality Theorem 4 is then proved
in the same way as Theorem 1, but this time there is no need to
distinguish between $L=0\bmod 4$ and $L=2\bmod 4$ --- or even
to restrict ourselves to one dimension.
\QEDnogroup

\bgsection 4. ASYMPTOTIC DIMERIZATION FOR $L=0\bmod 4$;\hfill\break
PROOFS OF THEOREMS 2 AND 3

Even though finite rings of length $L=0\bmod 4$ can do more
complicated things than dimerize
(see Section 5 for a discussion of counterexamples), they do
dimerize asymptotically. To be precise, for a fixed $f$, satisfying
conditions i) and ii) of \eq(conditions),
for which the infinite ring has a unique energy
minimizing configuration among the configurations of period 2, any
sequence of minimizers for finite rings has to approach that dimerized
configuration uniformily when $L\to\infty$, i.e., any nearest pair of
$\vert t_j\vert$'s  converges to the pair occurring in the infinite
volume dimerized configuration. The aim of this section is to prove
this statement, which is \Thm/asymptotics/ i). At the same
time we will also supply the proofs of
\Thm/uniquedim/ and \Thm/asymptotics/ ii).

In contrast to the previous section, here we will always consider
minimization of the energy over
arbitrary configurations  of $\vert t_j\vert$, but with constant
flux $\Phi$.
For this purpose it is convenient to introduce an explicit
notation for the energy functionals at fixed $\Phi$:
$$
\ee^\Phi_L(\{t_j\})=\ee_L(\{e^{i\Phi/L} t_j\})\quad .
$$
Here, and in the rest of this section, it is assumed that all $t_j$
are nonnegative. Note that the $t_j$'s  now play the role of the
$\vert t_j\vert$'s.

For any $x$ and $y$ let $d(x,y)$ denote the configuration with $t_j =
x$ for $j$ odd and $t_j=y$ for $j$ even, and define the energy density
$e$ of a period 2 configuration $d(x,y)$ as in \eq(density).
Note that $e$ is independent of $\Phi$.
In order to remove the degeneracy due to the obvious symmetry under
cyclic permutations  of the $t_j$'s, we will henceforth adopt the
convention that the maximum value of all $t$'s is attained by $t_1$.

A first ingredient in the proof of \Thm/asymptotics/ is the following
energy estimate, which by itself implies \Thm/asymptotics/ ii).
Let $E_L^\Phi$ denote the value of $\ee^\Phi_L$ in a minimizing
configuration $t^\Phi_{L,j}$. Under the general condition
\eq(conditions) ii) on
$f$ we can find a constant $K$, depending only on $f$, such that $\vert
t^\Phi_{L,j}\vert\leq  K$.

\iproclaim/energyestimate/ Lemma.
Assuming only condition ii) on $f$ in \eq(conditions), we have
for all $-\pi <\Phi \leq \pi$
$$\cases{
\vert E^0_L- E^\Phi_L\vert \leq 2\Phi^2 K/L
& if $L=2\bmod 4$\cr
\phantom{a} & \cr
\vert E^\pi_L- E^\Phi_L\vert \leq 2(\pi-\Phi)^2 K/L
& if $L=0\bmod 4$\cr
}$$
and
$$\cases{
\vert E^\pi_L - \ee^\Phi_L(t^0_L)\vert  \leq 4\pi^2 K/ L
& if $L=0\bmod 4$\cr
\phantom{a} & \cr
\vert E^\Phi_L - \ee^\Phi_L(t^\Psi_L)\vert  \leq 8\pi^2 K/ L
& for all $\Psi$ and all even $L$.\cr
}$$
\eproclaim

Because $\ee^\pi_L$ attains its minimum $E^\pi_L$ in a dimerized
configuration  ($L=0\bmod 4$), this lemma implies, for $\Phi=0$, that
even if the true minimum is not dimerized, its energy differs from the
best dimerized configuration by $\Order(1/L)$, which is 2 orders down
from $E^0_L=\Order(L)$.

\proof:
First note that $H(\{t_j\})$ has real matrix elements and therefore
has a
ground state  $\psi^0_L$ which is {\it real}, and hence such that
$\langle \psi^0_L\mid  \cd_{j+1,\sigma}
\c_{j,\sigma}\mid\psi^0_L\rangle$ is also real. Using a gauge
transformation that makes $H(\{\exp(i\pi/L)t_j\})$ real one can
also see that $H(\{\exp(i\pi/L)t_j\})$ has a
ground state $\psi^\pi_L$ such that
$$
\langle \psi^\pi_L\mid
\cd_{j+1,\sigma} \c_{j,\sigma}\mid\psi^\pi_L\rangle
= e^{i\pi/L} R_{j,\sigma}\quad ,
$$
with {\it real} $R_{j,\sigma}$.

Let $\{t^0_{L,j}\}$ be any minimizing configuration of $\ee^0_L$, and
let $\psi^0_L$ be a real ground state of $H(\{t^0_{L,j}\})$.
Then, by the variational principle,
$$\eqalign{
E^\Phi_L &\leq \langle \psi^0_L\mid H(\{\exp(i\Phi/L)t^0_{L,j}\})
\mid\psi^0_L\rangle + \sum_j f(t_j^0)\cr
&=\langle \psi^0_L\mid  H(\{\exp(i\Phi/L)t^0_{L,j}\})
\mid\psi^0_L\rangle + \sum_j f(t_j^0)\cr
&\quad +\sum_{j=1}^L\sum _{\sigma=\up,\down} (e^{i\Phi/L} -1)t^0_{L,j}
\langle \psi^0_L\mid \cd_{j+1,\sigma}\c_{j,\sigma} \mid\psi^0_L\rangle
+\quad \hc \quad .\cr
}$$
The sum of the first two terms in the right side of the equality
above  is $E^0_L$. Using the reality of $\psi^0_L$ the last term
can  be estimated by $4\vert 1-\cos ( \Phi/L)\vert\sum_j \vert
t^0_{L,j}\vert$.
As $\cos x\geq 1-x^2/2$, this proves the first estimate of the Lemma.
The other inequalities are derived in the same way.
\QED

The next lemma is an application of a version of
the  {\it Abstract Chessboard Estimate} \cite{FILS} which
is given in the Appendix.

\iproclaim/compareaveragedim/ Lemma.
For any configuration of $t_j\geq 0$, with $d(t_j,t_{j+1})$ described
in \eq(density),
$$\eqalign{
{1\over L}\sum_{j=1}^{L} \ee^0_L(d(t_j,t_{j+1}))
\leq \ee^0_L(\{t_j\}) & \quad\hbox{if } L=2\bmod 4\quad ,\cr
{1\over L}\sum_{j=1}^{L} \ee^\pi_L(d(t_j,t_{j+1}))
\leq \ee^\pi_L(\{t_j\}) &\quad\hbox{if } L=0\bmod 4\quad .\cr
}\deqno(compareaveragedim)$$
\eproclaim
\proof:
Let $\Phi=0$ if $L=2\bmod 4$ and $\Phi=\pi$ if $L=0\bmod 4$.
We apply \Thm15/chessboardestimate/ (with $D=\Rl^2$) to the
following $F$:
$$
F(a_1,\dots,a_{L})=\exp(-\ee^\Phi_L (t_1, \ldots,t_{L}))\quad ,
$$
where $a_j=(t_j,s_j)\in\Rl^2$. The involution $R$ is defined by
$R(t,s)=(s,t)$. The domain $\D$ is taken to be the set of all
$(a_1,a_2,\ldots,a_{L})$ of the form
$((t_1,t_2),(t_2,t_3),\ldots,(t_{L},t_1))$. The necessary invariance
properties of $\D$ are trivially satisfied. The cylicity of $F$ follows
from the invariance of $\ee^\Phi_L$ under cyclic permutations of the
$t_j$. The inequality of \Lem/RPineq/ translates
into Schwarz's inequality for $F$ on $\D$. This becomes obvious
with the definitions:
$$\eqalign{
a_1=(t_1,t_2) &\quad b_1=(t_{L},t_1)\cr
a_2=(t_2,t_3) &\quad b_2=(t_{L-1},t_{L})\cr
\vdots &\quad \vdots\cr
a_{L/2}=(t_{L/2},t_{L/2+1}) &\quad b_{L/2}=(t_{L/2+1},t_{L/2+2})
\quad .\cr
}$$
Indeed, one then has:
$$\eqalign{
\exp(-\ee^\Phi_L(t_1,\ldots,t_{L}))
&=F(a_1,a_2,\ldots,a_{L/2},b_{L/2},\ldots, b_1)\cr
\exp(-\ee^\Phi_L(t_1,\ldots,t_{L/2},t_{L/2+1},t_{L/2},\ldots,t_2))
&=F(a_1,a_2,\ldots,a_{L/2},R a_{L/2},\ldots, R a_1)\cr
\exp(-\ee^\Phi_L(t_1,t_{L},\ldots,t_{L/2+2},t_{L/2+1},\ldots,t_{L}))
&=F(R b_1, R b_2,\ldots, R b_{L/2},b_{L/2},\ldots, b_1)\cr
\exp(-\ee^\Phi_L(t_{L},t_{L-1},\ldots,t_{L/2+1},t_{L/2+2},
\ldots,t_1))
&=F(b_1, b_2,\ldots, b_{L/2},R b_{L/2},\ldots,R b_1)\cr
}$$
The implication of \Thm15/chessboardestimate/, in terms of
$\ee^\Phi_L$ then reads:
$$
\ee^\Phi_L(t)
\geq{1\over L}\sum_{j=1}^{L}
\ee^\Phi_L(t_j,t_{j+1},t_j,t_{j+1},\ldots)\quad ,
$$
which is the desired inequality.
\QED

\iproclaim/comparedim/ Lemma.
For the given $f$ let $K$ be a constant such that  $\vert
t^\Phi_{L,j}\vert\leq K$, for all $L$. For $L=0\bmod 4$,
let $\{t^0_{L,j}\}$ be any  minimizing configuration of $\ee^0_L$.
Then
$$
\vert\ee^\pi_L(d(t^0_{L,j},t^0_{L,j+1}))-E^\pi_L\vert
\leq 4K \pi^2\quad ,
$$
for all $j=1,2,\ldots, L$.
\eproclaim
\proof:
As $L=0\bmod 4$ we can apply \Lem/compareaveragedim/ to $\ee^\pi_L$
and  with $t_j=t^0_{L,j}$. We then have
$$
{1\over L}\sum_{j=1}^{L}\ee^\pi_L(d(t^0_{L,j},t_{L,j+1}))
\leq \ee^\pi_L(t^0_L)\quad .
$$
By \Lem/energyestimate/  the right side is bounded by
$E^\pi_L+4\pi^2 K/L$, and hence
$$
{1\over L}\sum_{j=1}^{L}\left\{\ee^\pi_L(d(t^0_{L,j},t^0_{L,j+1}))
-E^\pi_L\right\}\leq {4\pi^2 K \over L}\quad.
$$
As each of the terms in the sum is non-negative, the statement of the
lemma follows.
\QED

\iproclaim/energydensity/ Lemma.
i) The following
limit defines a continuous function $e$:
$$
\lim_{L\to\infty} \ee^\Phi_L(d(x,y))=e(x,y)
$$
The convergence is uniform on compact subsets of $\Rl^2$.\nl
ii) If $e$ attains its minimum at $(x_0,y_0)$ and $(y_0,x_0)$
and nowhere else, then  there exists a sequences $x_L\to x_0$ and
$y_L\to y_0$ such that for all $L=0\bmod 4$, $d(x_L,y_L)$ is a
minimizing configuration for $\ee^\pi_L$, and such that
$$
\lim_{L\to\infty}{1\over L}\ee^\pi_L(d(x_L,y_L))=e(x_0,y_0)\quad .
$$
\eproclaim
\proof:
The arguments involved in proving this lemma are fairly standard.
We therefore only give a sketch of the proof.

i) We first show that there exists a concave function
$\epsilon(x,y)$, satisfying  $\epsilon(x,y)\leq 2\vert
x\vert + 2\vert y\vert +\vert U\vert$, and such that
$$
\epsilon(x,y)=\lim_{L\to\infty}{1\over L} \lambda_0(H_L(d(x,y)))
\quad .
$$
For brevity we put $G_L= \lambda_0(H_L(d(x,y)))$. By the variational
principle and the fact that $\Vert c_{j\sigma}\Vert =1$ one
immediately gets that $G_L$ is ``almost subadditive'':
$$
G_{L_1+L_2}\leq G_{L_1} + G_{L_2} + 6(\vert x\vert +\vert y\vert)
$$
and therefore $G_L + 6(\vert x\vert +\vert y\vert)$ is subadditive.
This sequence is also almost monotone:
$$
L_1\geq L_2 \Rightarrow G_{L_1}\leq G_{L_2} + \hbox{Constant}
$$
because $G_L\leq 0$. It follows that the limit of $G_L/L$ exists and
as a limit of concave functions the limit is also concave. The bounds
are trivial but they imply continuity and therefore the convergence is
uniform on compacts. From \Lem/energyestimate/ it is obvious that
$e$ does not depend on $\Phi$.

ii) From the conditions on $f$ \eq(conditions) one has an apriori
bound $K$ on the
minimizing $t_j$: $\vert t^\pi_{j,L}\vert \leq K$. By \Thm/global/,
$\ee^\pi_L$ always has a dimerized minimizer $d(x_L,y_L)$, and by
compactness there always exists a subsequence $(x_{L_k},
y_{L_k})$ converging to $(x_*,y_*)$. By the continuity of $e^\pi$ and
the uniform convergence on compacts
$$
\left\vert {1\over L_k}\ee^\pi_{L_k}(d(x_{L_k},y_{L_k}))-e(x_*,y_*)
\right\vert\to 0 \quad .
$$
As the $d(x_{L_k},y_{L_k})$ are minimizers it follows that
$e(x_*,y_*)=e(x_0,y_0)$ and hence, by assumption,
$(x_*,y_*)=(x_0,y_0)$ or $(x_*,y_*)=(y_0,x_0)$. Without loss of
generality we can assume $x_0\geq y_0$. Consider then the new sequence
$d(\tilde x_L,\tilde y_L)$ of minimizers with $(\tilde x_L, \tilde
y_L)=(x_L,y_L)$ or $(y_L,x_L)$, and such that $\tilde x_L\geq \tilde
y_L$ for all $L$. Then all convergent subsequences $(\tilde x_{L_k},
\tilde y_{L_k})$ of $(\tilde x_L, \tilde y_L)$ converge to the same
limit $(x_0,y_0)$ and hence the sequence is convergent.
\QED

\noindent
{\bf Proof of \Thm/asymptotics/ i):}\nl
Let $d(x_L,y_L)$ be a sequence of minimizing configurations
for $\ee^\pi_L$ and such that $x_L$ and $y_L$ converge to $x_0$ and
$y_0$ respectively as $L\to\infty$. The existence of such a sequence
is guaranteed by \Lem/energydensity/.

First we estimate $e(t^0_{L,j},t^0_{L,j+1})$ for a minimizing
configuration $t^0_L$ of $\ee^0_L$ and any \break
$j=1,\ldots, L$.
$$\eqalign{
&\vert e(t^0_{L,j},t^0_{L,j+1})-e(x_0,y_0)\vert\cr
&\leq \left\vert e(t^0_{L,j},t^0_{L,j+1})-{1\over L}
\ee^\pi_L(d(t^0_{L,j},t^0_{L,j+1}))\right\vert\cr
&\quad + {1\over L}\left\vert\ee^\pi_L(d(t^0_{L,j},t^0_{L,j+1}))
-\ee^\pi_L(d(x_L,y_L))\right\vert
+{1\over L}\left\vert\ee^\pi_L(d(x_L,y_L)) -e(x_0,y_0)\right
\vert \quad .\cr
}$$
All three terms on the right side vanish as $L\to\infty$, uniformly in
$j$: the first term by the uniform convergence on compact sets  given
by \Lem/energydensity/ i), the second by \Lem/comparedim/, and the
third term by  \Lem/energydensity/ ii).

The theorem now follows from the uniqueness of the dimerized minimum
and compactness. Indeed, suppose one can find arbitrarily large $L_k$
such that
$$
\vert t^0_{L_k,j}-x\vert +\vert t^0_{L_k,j+1} - y\vert \geq \epsilon
\quad \hbox{ and } \quad \vert t^0_{L_k,j}-y\vert +
\vert t^0_{L_k,j+1} - x\vert\geq \epsilon \quad ,
$$ for some $\epsilon >0$.
By compactness and the continuity of $e$ there would then be a
limit point $(x^\prime, y^\prime)$ of $(t^0_{L_k, j},
t^0_{L_k, j+1})$,
for which $e(x^\prime,y^\prime)=e(x_0,y_0)$.
As $(x^\prime,y^\prime)$ as at least distance $\epsilon$ away from
$(x_0,y_0)$
this
would imply the existence of another minimizer for $e$ which is
ruled out by  assumption.
\QEDnogroup

\noindent
{\bf Proof of \Thm/uniquedim/ :}\nl
Let $\Phi=0$ if $L=2\bmod 4$ and $\Phi=\pi$ if
$L=0\bmod 4$. Denote by $d(x,y)$ the unique dimerized
minimizer of $\ee^\Phi_L$. Let $\{t^\Phi_j\}$ be any
minimzer of $\ee^\Phi_L$. As $\ee^\Phi_L$ attains its minimum
in $\{t^\Phi_L\}$, the inequality of \Lem/compareaveragedim/
is an equality for $t_j=t^\Phi_j$. Therefore, all configurations
on the left side of \eq(compareaveragedim) must be minimizers.
These configurations are all dimerized and therefore (by our
uniqueness assumption) must all be equal to $d(x,y)$.
\QEDnogroup

\bgsection 5. COUNTEREXAMPLES: RINGS OF LENGTH $L=0\bmod 4$\hfill\break
HAVING MINIMIZERS AT ZERO FLUX THAT ARE NOT PERIOD TWO

We will only discuss counterexamples with $U=0$. It is obvious, then,
that counterexamples also exist with small non-vanishing $U$.
However, the presence of a not too large on-site repulsion  tends to
enhance the dimerization, while nearest neighbour repulsions can
reduce it. We have not investigated the effects of interactions on the
new instabilities discussed in this section.  Our result for
the spin-Peierls problem, \Thm/spinpeierls/, which can be interpreted
as the $U\to\infty$
limit \cite{And}, seems to indicate that, when $U$ is very large,
the translation  symmetry can
break down from period one to at most period two. This seems
natural because the electrons are more mobile when
$U$ is small and, under this condition,  other
deformations of the lattice can become favorable if the flux is  not
optimal. We interpret distortions that break the periodicity two as an
attempt of the system to mimic the energy lowering effect  of a
nonvanishing flux. This is also a possible interpretation of the
lattice distortions found in two dimensions \cite{Maz,TH}.

Kennedy and Lieb \cite{KL} gave an example of a function $f$
(involving a quartic term) which gives rise to a minimum for the model
on a square ($L=4$) which does {\it not\/} have periodicity two.
Longuet-Higgins and Salem pointed out the possibility of other than
period two instabilities a long time ago \cite{LHS}, but they expected
these to be unimportant and thought the period two instability
would always  dominate. They only considered the case $L=2\bmod 4$
and therefore did not notice that other
instabilities can, in fact, occur when $L=0\bmod 4$.

Our aim in this section is twofold. First we present a family of
functions $f$ that lead to non-period-two minimizers of the
zero-flux problem for the ring of
eight sites. Then we discuss what instabilities other than period two
occur. From the discussion it will become clear how to construct
many more examples of non-period-two minima. The properties of the new
instabilities, which we study partly only numerically, support our
conjectures stated in the introduction.

\medskip
\noindent
{\bf Counterexample for $L=8$.}\nl
For simplicity we will present just one family of functions  $f$
chosen to yield simple values for the minimizing  dimerized
configuration. It is easy to generalize this example in several
directions.

Define a function $f(t)$, for $t\geq 0$, as follows (see Figure 2 for
the graph of $f$.)
$$
f(t)=\cases{
{1\over 10} + {1\over 5} (t-1) +{1\over 10} (1-t)^2
&if $0\leq 4-a$\cr
\phantom{a} & \cr
{1\over 10} + {1\over 5} (t-1)+{1\over 10}(1-t)^2
+{1\over 3a^2}(t-4+a)^3
&if $t\geq 4-a$\quad .\cr
}\deqno(f)$$
Here $a$ can be any number strictly between $0$ and $4-5/\sqrt{2}\cong
.4645$.

Consider first configurations $t_j$ of the form $t_{2j+1}=x$,
$t_{2j}=y$. We always assume $x\geq y\geq 0$. The ground state energy
of eight electrons on a ring of eight sites described by the
Hamiltonian \eq(hubham) with $U=0$ is then $4 e(x,y)$ where $e(x,y)$
is given by
$$
e(x,y)= -x-\sqrt{x^2+y^2}\quad .
$$
For the 8-ring with $f$ given in \eq(f), the optimum period two
configuration is then determined by the values of $x_0\geq y_0$ that
minimize
$$
h(x,y)\equiv e(x,y) + f(x) + f(y) \quad .
$$
It is trivial to see that $x_0>0$, and as $h$ is differentiable away
from $x=0$ and $h(x,y)\to +\infty$ as $x$ or $y\to +\infty$,
$(x_0,y_0)$ must therefore be a critical point of $h(x,y)$ solving the
equations
$$\eqalign{
1+ {x\over \sqrt{x^2+y^2}}&=f^\prime(x)\cr
{y\over \sqrt{x^2+y^2}}&=f^\prime(y)\quad .\cr
}$$
A straightforward calculation shows that $(x,y)=(4,3)$ is the unique
minimum.

Next we study the stability of the relative minimum $(x_0,y_0)$ under
a class of perturbations which break the period two symmetry. For
future use we do this for arbitrary $L=0\bmod 4$, $L\geq 8$. More
specifically we consider configurations of the form $t_{2j+1}=x$,
$t_{2j}=y+\epsilon V_{j+1}$, $j=1,\ldots, N\bmod N$, with $L=2N$. As
we do not want to introduce a flux, the $V_j$ must be real. One can
use analytic perturbation theory to compute the  energy as a function
of $x,y$ and $\epsilon$. Up to corrections of $\Order(\epsilon^3)$ one
finds
$$\eqalign{
\hbox{Kin. Energy } =&-2\sum_{n=1}^N \lambda_n -
\epsilon \hat V (0)\sum_{n=1}^N {y+x\cos(\alpha n)\over \lambda_n}\cr
&+\epsilon^2\sum_{n=1}^N\left({\vert\hat V (0)\vert^2
(y+x\cos(\alpha n))^2\over \lambda_n^3}
-{\sum_{m\neq 2n\bmod N} \vert\hat V(m)\vert^2
\over \lambda_n} \right)
-\epsilon^2{\vert \hat V(0)\vert^2 \over \vert x-y\vert}\cr
&-4\epsilon^2 \sum_{n,m;\lambda_n \neq\lambda_m}
\vert \hat V(n-m)\vert^2
{P_{n,m} \over \lambda_n(\lambda_n^2-\lambda_m^2)} \quad ,\cr
}\deqno(delta)$$
where
$$
\lambda_n=\sqrt{x^2+y^2+2xy\cos(\alpha  n)},\quad \alpha=2\pi/N \quad .
$$
and
$$
P_{n,m}= y^2 + x^2 \cos(\alpha(n+m)/2))^2 +2xy \cos(\alpha(n+m)/2)
\cos(\alpha(n-m)/2) \quad .
$$
$\hat V$ is the Fourier transform of $V_j$: $\hat V(q)= (1/N)\sum_j
\exp (2\pi ijq/N) V_j$. Note that the coeffient of $\epsilon^2$ is
diagonal in $\hat V$. Up to terms $\Order(\epsilon^3)$, the elastic
energy is
$$
N\left( f(y) +f(x) + \epsilon f^\prime(y)\sum_j V_j+
\tover12 \epsilon^2
f^{\prime\prime}(y)\sum_j V_j^2\right) \quad .
$$
In a critical point $(x,y)$ in the set of period two  configurations,
the terms proportional to $\epsilon$ in the electron energy will
cancel the ones in the elastic energy. A perturbation $V$ will then
lower the energy for small $\epsilon$ if the total (electron plus
elastic energy) coefficient of  $\epsilon^2$ is negative, which means
that the optimal dimerized configuration is not the true minimum. Call
the coefficient of the $\epsilon^2$ contribution to the electron
energy $-\Delta(x,y)$. Note that this coefficient is always negative.
There is an instability whenever
$$
\Delta(x_0,y_0) > {N\over 2}f^{\prime\prime}(y_0) \quad .
\deqno(instab)$$

We now return to the model on the 8-ring and show that
indeed the energy can be lowered by perturbing the dimerized minimum
$(x_0,y_0)$ found above. For the 8-ring ($N=4$) there are four
independent perturbations  $V_i$ to consider:
$$\eqalign{
V=(1,1,1,1),&\quad V=\sqrt{2}(1,0,-1,0)\quad ,\cr
V=(1,-1,1,-1),&\quad V=\sqrt{2}(0,1,0,-1)\quad .\cr
}$$
The first, $V=(1,1,1,1)$, respects the period 2 and in the optimum
dimerized configuration it cannot lower the energy. The second and the
fourth have the same energy because they are translates of each other.
We have normalized the $V$'s such that $\sum_i V_i^2=4$ (in general we
will use the normalization $\sum_i V_i^2=N$.) Let
$-\Delta_0,-\Delta_1,-\Delta_2$ be the coefficients of $\epsilon^2$
for the first, second and third choice of $V$ respectively. It is
convenient to factor out $x$ as a scale factor and to use the ratio
$r=y/x$ as the  relevant variable, $0\leq r\leq 1$. The explicit
expressions for the $\Delta$'s are:
$$\eqalign{
x\Delta_0&={2\over (1+r^2)^{3/2}}\cr
x\Delta_1&= 4 - {2\over \sqrt{1+r^2}}\cr
x\Delta_2&=2 \quad .\cr
}$$
In our example both $\Delta_1$ and $\Delta_2$ satisfy the criterion
for instability \eq(instab):
$$
\vert \Delta_1(4,3)\vert
= {3\over 5}>2f^{\prime\prime}(3)= {2\over 5},\qquad
\vert \Delta_2(4,3)\vert
= {1\over 2}>2f^{\prime\prime}(3)= {2\over 5} \quad .
$$
Of course, $\Delta_0(4,3)$ does not satisfy \eq(instab), indicating
that (4,3) is a  stable relative minimum within the set of period two
configurations (indeed $\Delta_0(4,3)=32/125 < 2/5$). We conclude that
for the choice of $f$ defined in \eq(f) the energy minimizing
configuration of the 8-ring is not period two.

\medskip
\noindent
{\bf Other than period-two instabilities for arbitrary $L
=0\bmod 4$.\/}\nl
The behavior of the coefficients of $\epsilon^2$ in \eq(delta) for
large $N$ gives support to our conjectures 1 and 2
stated in the introduction.

Let $-\Delta_k$ denote the coefficient of $\epsilon^2$ for  $V=
V^{(k)}(q)$, where $\hat
V^{(k)}(q)=(\delta_{q,k}+\delta_{q,N-k})/\sqrt{2}$, $k\neq 0,N/2$,
$\hat V^{(0)}(q)=\delta_{q,0}$, $\hat V^{(N/2)}(q)=\delta_{q,N/2}$.
The perturbations are normalized such that their contributions to the
elastic energy is identical. $\Delta_0$
and $\Delta_1$ are  given by:
$$\eqalign{
\Delta_0&= \sum_{n=1}^N {x^2\sin(\alpha n)^2\over
\lambda_n^{3}} \quad ,\cr
\Delta_1&= \sum_{n=1}^N {1\over\lambda_n}+
\sum_{n=1}^N\left({1\over \lambda_n}-{1\over\lambda_{n+1}}\right)
{P_{n,n+1}\over xy (\cos(\alpha n) -\cos(\alpha (n+1)))}\quad .\cr
}$$
The stability of a relative minimum $(x_0,y_0)$ within the set of
period two configurations requires that $\Delta_0 <
(N/2)f^{\prime\prime}(y_0)$. When $\Delta_0 <
(N/2)f^{\prime\prime}(y_0)< \Delta_1$ such a relative minimum is
unstable against the perturbation $\epsilon V^{(1)}$, which has
wavelength $=L$. Numerical investigation of these expressions reveals
the following: for $r\equiv y/x <1$ but close enough to 1 (depending
on $N$), one has $\Delta_1 >\Delta_0$. This is illustrated in
Figure 3. Note  that the perturbations with $k >1$ are all stable. We
showed in the previous subsection that for $L=8$ this situation can be
realized with a function $f$ satisfying the conditions i and ii) stated
in the introduction. It is reasonable to expect that this can be done
for all $L$ that are a multiple of 4, but the larger $L$ becomes the
more stringent are the restrictions on $f$. In particular, if one fixes
$f$ we expect that for large enough systems the long wavelength
instability will not occur and the system will remain in a period two
configuration.

\bgsection A. Appendix

\noindent
{\bf 1) Discontinuous and non-convex $f$\/}\nl
We have assumed throughout this paper that $f$ was a continuous
function satisfying some simple conditions as stated in the
introduction \eq(conditions).
Whether or not $f$ is continuous we can associate
a function $\overline f$ with $f$ having the following
properties: (a) $\overline f$ is convex (i.e., $\overline f(\lambda t
+(1-\lambda)t^\prime) \leq \lambda \overline f(t) +
(1-\lambda)\overline f(t^\prime)$ for all $0\leq \lambda \leq 1$ and
all $t,t^\prime \geq 0$; (b) $\overline f(t)
=\sup\{g(t)\, :\, g \hbox{ is convex and }
g(t^\prime)\leq f(t^\prime) \hbox{ for all } t^\prime\geq 0\}$. This
function $\overline f$ is called the convex hull of $f$ and is
automatically continuous for $t>0$. The following lemma shows
that all we really need is that $\overline f$ (not $f$)
satisfy conditions i) and ii) of \eq(conditions).

\iproclaim/convexf/ Lemma.
Let $\{t_j^\Phi\}$ be a minimizing configuration of $\ee^\Phi_L$.
Then for all $j$, $f(t^\Phi_j)=\overline f (t^\Phi_j)$, where
$\overline f$ is the convex hull of $f$.
\eproclaim
\proof:
Let $j$ be arbitrary and fixed. As a function of $t_j=t$ the
energy can be written as
$$\eqalign{
\ee(t)&=E(t) + f(t)\cr
&=(E(t) + at) + (f(t)-at)=K(t) + g(t) \quad .\cr
}$$
We choose $a = (f(t_2)-f(t_1))/(t_2-t_1)$ where
$$
t_1=\sup\{s<t_j^\Phi\mid f(s)\neq \overline{f} (s)\},\quad
t_2=\inf\{s>t_j^\Phi\mid f(s)\neq \overline{f} (s)\} \quad .
$$
As $t$ appears linearly in the Hamiltonian $H$, $K(t)=E(t) + at$ is a
concave function of $t$ (i.e., $-E(t)-at$ is convex). By construction
we also have that $g(t_j^\Phi)\geq g(t_1)=g(t_2)$.
Hence, with $0\leq\alpha\leq 1$ such that
$t_j^\Phi=\alpha t_1 + (1-\alpha) t_2$ one has
$$
\ee(t_j^\Phi)\geq \alpha \ee(t_1)
+(1-\alpha)\ee(t_2)
$$
and the inequality is strict if $g(t_j^\Phi)=f(t_j^\Phi)
-\overline f (t_j^\Phi)\neq 0$. Therefore the minimium can be attained
only at $t_j$ where $f(t_j)= \overline f (t_j)$.
\QED

\medskip
\noindent
{\bf 2) The DLS Lemma for ground states}\nl
The following lemma is a ground state version of Lemma 4.1 in
\cite{DLS}. An example of this ground state version is given in
\cite{KLS}. Note that the matrices $A$ and $B$ need not have real
matrix elements. It is obvious how to generalize to reflections  that
involve an additional unitary transformation. When needed one can in
fact apply the lemma as stated below after performing a unitary
transformation on one half of the system. In this paper we applied
the lemma in two different situations. In the proof of \Lem/spin0/
the two copies of the Hilbert space $\H$ correspond to spin $\up$ and
spin $\down$. In \Lem/RPineq/ the two spaces refer to the right and
left halves of the ring.

\iproclaim/DLS/ Lemma.
Let $A,B,C_1,\ldots,C_n$ be a collection of $d\times d$ complex
matrices ($n$ could be infinite) with the following properties:
$A$ and $B$ are Hermitian (i.e., $A=A^\dagger, B=B^\dagger$),
for all $i$,
$C_i$ is real and $\sum_i C_i\otimes C_i$
is symmetric (as a $d^2\times d^2$ matrix). Let $\lambda_0(A,B)$
denote the lowest eigenvalue of the matrix
$$
T(A,B)\equiv A\otimes\idty-\sum_i C_i\otimes C_i
+\idty\otimes B \quad .
\deqno(TAB)$$
Then
$$
\lambda_0(A,B)\geq \tover12\left( \lambda_0(A,A^*
)+\lambda_0(B^* ,B)\right) \quad , \deqno(TABineq)$$ where
$A^*$ denotes the matrix obtained from $A$ by complex
conjugation of the matrix elements. In particular
$$
\lambda_0(A,B)\geq \min\left(
\lambda_0(A,A^* ),\lambda_0(B^* ,B)\right) \quad .
$$
The inequality \eq(TABineq) is strict if for some $i$
$$
\langle\Omega\mid C_i\otimes\idty\mid\Omega\rangle \neq
\langle\Omega\mid
\idty\otimes C_i\mid\Omega\rangle \quad , \deqno(strict)$$ where
$\Omega$
is any eigenvector for the eigenvalue $\lambda_0(A,B)$ (this is not
the only case where the inequality is strict.) In particular all
ground states $\Omega$ of $T(A,A^*)$ satisfy
$$
\langle\Omega\mid C_i\otimes\idty\mid\Omega\rangle
= \langle\Omega\mid \idty\otimes C_i\mid\Omega\rangle \quad .
$$
\eproclaim
\proof:
Let $\{e_j\}$ be a orthonormal basis of $\Cx^d$ in which the $C_i$
have real matrix elements. Define $V$ to be the antilinear map
leaving the $e_j$ invariant: $Ve_j=e_j$. Then $VAV=A^*$ etc.
Let $\Omega$ be an eigenvector of the smallest eigenvalue of the
operator \eq(TAB) and denote by $M_{ij}$ its coefficients in the basis
$e_i\otimes e_j$:
$$
\Omega=\sum_{i,j} M_{ij}e_i\otimes e_j \quad .
$$
We can consider $(M_{ij})$ as a matrix and use the eigenvectors
$\{u_\alpha\}$ of $M^\dagger M$ and $\{v_\alpha\}$ of $MM^\dagger$
to define two new bases:
$$\phi_\alpha=\sum_j \langle e_j\mid u_\alpha\rangle e_j
\quad ,\qquad \psi_\alpha=\sum_j \langle e_j
\mid v_\alpha\rangle^* e_j \quad .
$$
It is then straigthforward to check that
$$
\Omega=\sum_\alpha \lambda_\alpha \phi_\alpha\otimes\psi_\alpha \quad .
$$
We can obviously assume that $\lambda_\alpha >0$ by absorbing a phase
in the definition of $\phi_\alpha$ and by omitting any vanishing terms
in the sum.

Using the fact that $\{\phi_\alpha\}$ and $\{\psi_\beta\}$ are
orthonormal bases the energy of $\Omega$ can be expressed as
$$\eqalign{
\langle\Omega\mid T(A,B)\mid\Omega\rangle
&= \sum_\alpha \lambda_\alpha^2\left(\langle\phi_\alpha\mid
A\mid\phi_\alpha\rangle
+\langle\psi_\alpha\mid B\mid\psi_\alpha\rangle\right)\cr
&\quad - \sum_i\sum_{\alpha\beta}\lambda_\alpha \lambda_\beta
\langle\phi_\alpha\mid C_i\mid\phi_\beta\rangle
\langle\psi_\alpha\mid C_i\mid\psi_\beta\rangle \quad .\cr
}$$
This energy can directly be compared with  the energy of
$T(A,A^*)$ and $T(B^* , B)$ in the states $\Omega_1$
and $\Omega_2$ defined by
$$\eqalign{
\Omega_1=\sum_\alpha \lambda_\alpha \phi_\alpha\otimes V\phi_\alpha\cr
\Omega_2=\sum_\alpha \lambda_\alpha V\psi_\alpha\otimes \psi_\alpha\cr
}$$
which have the same norm:
$\Vert\Omega\Vert^2=\Vert\Omega_1\Vert^2=\Vert\Omega_2\Vert^2=
\sum_\alpha \lambda_\alpha^2$. The average of the energies of
$\Omega_1$ and $\Omega_2$ is
$$\eqalign{
&\tover12(\langle\Omega_1\mid T(A,A^* )\mid\Omega_1\rangle
+ \langle\Omega_2\mid T(B^* , B)\mid\Omega_2\rangle)\cr
&= \tover12\sum_\alpha \lambda_\alpha^2
\left(\langle\phi_\alpha\mid A\mid\phi_\alpha\rangle
+\langle V\phi_\alpha\mid A^* V\mid\phi_\alpha\rangle
+\langle V\psi_\alpha\mid B^* V\mid\psi_\alpha\rangle
+\langle\psi_\alpha\mid B\mid\psi_\alpha\rangle\right)\cr
&- \tover12\sum_i\sum_{\alpha\beta}\lambda_\alpha \lambda_\beta
\left(\langle\phi_\alpha\mid C_i\mid\phi_\beta\rangle
\langle V\phi_\alpha\mid C_i V\mid\phi_\beta\rangle
\quad +\langle \psi_\alpha\mid C_i\mid\psi_\beta\rangle
\langle V\psi_\alpha\mid C_i V\mid\psi_\beta\rangle\right).\cr
} \deqno(**)$$
Using $\langle V\phi\mid \psi\rangle=\langle \phi\mid
V^\dagger\mid\psi\rangle^*$, for any two vectors $\phi,\psi$, the
hermiticity of $A$ and $B$ and the reality of the $C_i$, it is
straightforward to verify:
$$\eqalign{
\langle V\phi_\alpha\mid A^* V\mid\phi_\alpha\rangle
&=\langle \phi_\alpha\mid A \mid\phi_\alpha\rangle\cr
\langle V\phi_\alpha\mid C_i V\mid\phi_\beta\rangle
&=\langle \phi_\alpha\mid C_i \mid\phi_\beta\rangle^*\quad ,\cr
}$$
etc. As $\lambda_\alpha >0$ we can then apply $2\vert u v\vert\leq
\vert u\vert^2+\vert v\vert^2$ to each term in the double sum of
\eq(**) to obtain
$$
\langle\Omega\mid T(A,B)\mid\Omega\rangle
\geq \tover12\left(\langle\Omega\mid T(A,A^* )\mid\Omega\rangle
+\langle\Omega\mid T(B^* ,B)\mid\Omega\rangle\right)
$$
from which \eq(TABineq) directly follows by the variational principle.

The inequality is strict if for some $i$ and some $\alpha, \beta$,
$\langle \phi_\alpha\mid C_i \mid\phi_\beta\rangle \neq \langle
\psi_\alpha\mid C_i \mid\psi_\beta\rangle$, for which \eq(strict) is a
sufficient condition. When $A=B^*$, \eq(TABineq) is an equality
and so $\langle \phi_\alpha\mid C_i \mid\phi_\beta\rangle \neq \langle
\psi_\alpha\mid C_i \mid\psi_\beta\rangle$ for all $i,\alpha,\beta$ in
that case.
\QED

{}From the proof of the lemma it also follows that any Hamiltonian of
the form $T(A,A^*)$, has a ground state $\Omega$ satisfying
$$
\langle\Omega\mid X\otimes X^* \mid\Omega\rangle \geq 0 \quad .
$$
for all observables $X$ on $\H$.

The same state therefore also satisfies
$$
\langle\Omega\mid X\otimes \idty \mid\Omega\rangle
=\langle\Omega\mid \idty\otimes X^* \mid\Omega\rangle \quad .
$$
In particular if $[X\otimes\idty,T(A,A^* )]=0$, and $X=X^\dagger$,
there will be a ground state $\Omega$ such that $X\otimes \idty \Omega
=\idty\otimes X^* \Omega = x\Omega$ for some eigenvalue $x$ of
$X$. (This also follows from application of \eq(TABineq) with $A$
replaced by $A+ c(X-x_1)^2$ and $B$ by $A^* + c (X^*
-x_2)^2$, where $x_1$ and $x_2$ are the eigenvalues of $X\otimes\idty$
and $\idty\otimes X^*$ in some ground state of $T(A,A^*)$.

\bigskip
\noindent
{\bf 3) The Abstract Chessboard Estimate(\cite{FILS})\/}\nl
\iproclaim/chessboardestimate/ Theorem.
Let $D$ be a set,  let $R:D\to D$ be an involution on $D$ ($R^2=$
identity), and
let $F:\D\to\Cx$ be a complex valued function on a domain $\D\subset
D^{\times 2N}$ that has the following invariance properties:
$$\eqalign{
(a_1,\ldots,a_{2N})\in\D &\imply (a_2,\ldots,
a_{2N},a_1)\in\D \quad ,\cr (a_1,\ldots,a_N,b_N,\ldots,b_1)\in\D
&\imply (a_1,\ldots, a_N, R a_N, \ldots,R a_1),
(b_1,\ldots, b_N, R b_N,\ldots R b_1)\in\D
%\textbox{6}{.7}{ $(a_1,\ldots, a_N, R a_N, \ldots,R a_1)$ and
%\hfill\break $(b_1,\ldots, b_N, R b_N,\ldots R b_1)\in\D$\quad .}
\cr
}$$
Assume $F$ obeys
$$
F(a_1,\ldots,a_{2N})=F(a_2,\ldots,a_{2N},a_1)
$$
and
$$\vert Fa_1,\ldots,a_N,b_N,\ldots,b_1)\vert^2\leq F(a_1,\ldots, a_N,
R a_N, \ldots, R a_1) F(b_1,\ldots , b_N, R b_N,\ldots, R b_1) \quad .
$$
Then $\Vert a\Vert\equiv \vert F(a,R a, A,\ldots,R a)\vert^{1/2N}$
satisfies
$$
\vert F(a_1,\ldots,a_{2N})\vert
\leq \prod_{j=1}^{2N} \Vert a_i\Vert \quad .
$$
\eproclaim

The proof of the Theorem is exactly the same as in \cite{FILS} Theorem
4.1. In that reference it is also assumed that $D$ is a real vector
space and that $F$ is defined on all of $D^{2N}$. Under these
circumstances the inequality of the theorem implies that
$\Vert\,\cdot\, \Vert$ is a seminorm on $D$, but we do not use this
property in this paper.

\medskip
\noindent
{\bf ACKNOWLEDGEMENTS}

We gratefully acknowledge helpful discussions with
P. Wolff, J. Morgan, J.-Ph. Solovej, Z. Soos,
A. Moscowitz. Both authors were partially supported by the
U. S. National Science Foundation under  grant PHY90-19433 A03.

\bigskip

\noindent
{\bf REFERENCES}
\bigskip

\let\REF\doref
\parskip=0pt
\baselineskip=12pt
\refskip=8pt
%\refskip=12pt

\REF Hue Hue \Gref
E. H\"uckel
"Quantumtheoretische Beitr\"age zum Benzolproblem. I-III."
Z. Phys. {\bf 70}\hfill\break
 (1931) 204-292, {\bf 72} (1931) 310-337, {\bf 76} (1932) 628-648

\REF Sal Sal \Bref
L. Salem
"The Molecular Orbital Theory of Conjugated Systems"
W.A. Benjamin Inc., New York-Amsterdam 1966

\REF Lon Lon \Jref
F. London
"Th\'eorie quantique des courants interatomiques dans
les combinaisons aromatiques"
J. de Phys. Radium (now J. de Phys.),
S\'erie VII, Tome VIII @ (1937) 397-409

\REF Jon Jon \Jref
H. Jones
"Applications of the Bloch Theory to the Study
of Alloys and the Properties of Bismuth"
Proc. Roy. Soc. (London) @A147(1934)396-417

\REF Hub Hub \Jref
J. Hubbard
"Electron correlations in narrow energy bands"
Proc. Roy. Soc. (London) @A276(1963)238-257

\REF PP PP \Jref
R. Pariser, R.G. Parr
"A semi-empirical theory of the electronic spectra and
electronic structure of complex unsaturated hydrocarbons
I. and II"
J. Chem. Phys. @21(1953)466-471

\REF Pop Pop \Jref
J.A. Pople
"Electron interaction in unsaturated hydrocarbons"
Trans. Faraday Soc. @49(1953)1375-1385

\REF Lie4 Lie4 \Gref
E.H. Lieb
"The Hubbard model -- Some Rigorous results and open problems"
in Proceedings of 1993 conference in honor of G.F. Dell'Antonio, {\it
Advances in dynamical systems and quantum physics}, World Scientific (in
press) and in Proccedings of 1993 NATO ASW {\it The Physics and
mathematical physics of the Hubbard model}, Plenum (in press).

\REF SSH SSH \Jref
W.P. Su, J.R. Schrieffer, A.J. Heeger
"Soliton excitations in polyacetylene"
Phys. Rev. @B22(1980)2099-2111

\REF KH KH \Jref
S. Kivelson, D.E. Heim
"Hubbard versus Peierls and the Su-Schrieffer-Heeger model
of polyacetylene"
Phys. Rev. @B26(1982) 4278-4292

\REF BCM BCM \Gref
D. Baeriswyl, D.K. Campbell, S. Mazumdar
"An Overview of the Theory of $\pi-$Con\-jug\-ated
Polymers"
pp 7-133 \inPr H. Kiess
"Conjugated Conducting Polymers"
Springer Series in Solid State Sciences 102,
Springer, New York - Berlin, 1992

\REF BJ BJ \Gref
D. Baeriswyl, E. Jeckelmann
"The Hubbard model and its application
to conjugated $\pi$-electron systems"
preprint, to appear in Proceedings
San Sebastian

\REF Pei Pei \Bref
R.E. Peierls
"Quantum Theory of Solids"
Clarendon, Oxford 1955, p. 108

\REF Fro Fro \Jref
H. Fr\"ohlich
"On the theory of superconductivity: the
one-dimensional case"
Proc. Roy. Soc. (London) @A223(1954)296-305]

\REF LHS LHS \Jref
H.C. Longuet-Higgins, L. Salem
"The alternation of bond lengths in long conjugated
chain molecules"
Proc. Roy. Soc. (London) @A251(1959)172-185

\REF Lab Lab \Jref
H. Labhart
"FE Theory Including an Elastic $\sigma$ Skeleton.
I. Spectra and Bond Lengths in Long Polyenes"
J. Chem. Phys. @27(1957)957-962

\REF Oos1 Oos1 \Jref
Y. Ooshika
"A Semi-empirical Theory of the Conjugated Systems.
I. General Formulation"
J. of the Phys. Soc. of Japan @12(1957)1238-1245

\REF KL KL \Jref
T. Kennedy, E.H. Lieb
"Proof of the Peierls Instability in One Dimension"
Phys. Rev. Lett. @59(1987)1309-1312

We take the opportunity to note three minor technical misstatements
in this paper. (A.) The paragraph after eq.\ (6) is not correct for
$N=2$ because $T_{13}=t_1t_2 + t_3t_4$ and not $t_1t_2$ in this case.
Thus, $z$ has to be replaced by $2z$, but the rest of the argument
works. Alternatively, one can compute the eigenvalues of $T^2$
explicitly since it reduces to a $2\times 2$ matrix. (B.) The
uniqueness proof for $N=2$ needs strengthening. $T^2=\langle
T^2\rangle$
implies only that $t_1=t_3$ or $t_2=t_4$. However, this case can be
analyzed explicitly and uniqueness holds. (C.) The statement in {\it
Case 1\/} that $W(z)\equiv\Tr(2y^2 + z\Omega)^{1/2}$ is an even
function of $z$ is correct only when $N$ is even (because the
sublattices are then themselves bipartite). However, in all cases
$W(z)$ is certainly concave and
$d W(z)/dz=(1/2)\Tr(2y^2 + z\Omega)^{-1/2}\Omega$, which is zero at
$z=0$. Thus, $W(z)$ is decreasing for $z>0$ and increasing for $z<0$,
which is what is needed in {\it Case 1\/} and {\it Case 2\/}.
\vskip\refskip

\REF Imr Imr \Gref
Y. Imry
"Physics of mesoscopic systems"
pp 101-163
\inPr G. Grinstein, G. Mazen\-ko
"Directions in Condensed Matter Physics"
World Scientific, Singapore 1986

\REF Wie Wie \Jref
P.W. Wiegmann
"Towards a gauge theory of strongle correlated
electronic systems"
Physica @C153(1988)102-108

\REF Lie3 Lie3 \Jref
E.H. Lieb
"The flux phase problem on planar lattices"
Helv. Phys. Acta @65(1992)247-255

\REF LS LS \Jref
E.H. Lieb, M. Loss
"Fluxes, Laplacians, and Kasteleyn's theorem"
Duke Math. J. @71(1993)337-363

\REF Lie2 Lie2 \Jref
E.H. Lieb
"The Flux-Phase of the Half-Filled Band"
Phys. Rev. Lett. @73(1994)2158-2161

\REF Oos2 Oos2 \Jref
Y. Ooshika
"A Semi-empirical Theory of the Conjugated Systems.
II. Bond Alternation in Conjugated Chains"
J. of the Phys. Soc. of Japan @12(1957)1246-1250

\REF Gar Gar \Bref
P.J. Garratt
"Aromaticity"
McGraw-Hill, London 1971

\REF And And \Jref
P.W. Anderson
"New Approach to the Theory of Superexchange Iteractions"
Phys. Rev. @115(1959)2-13

\REF Chs Chs \Jref
D.B. Chesnut
"Instability of a Linear Spin Array: Application to W\"urster's Blue
Perchlorate"
J. Chem. Phys. @45(1966)4677-4681

\REF Pin Pin \Jref
P. Pincus
"Instability of the uniform antiferromagnetic chain"
Solid State Commun. @9(1971)1971-1973

\REF BP BP \Jref
G. Beni, P. Pincus
"Instability of the Uniform Antiferromagnetic Chain.
I. $XY$ Model in the Adiabatic Approximation"
J. Chem. Phys. @57(1972)3531-3534

\REF CF CF \Jref
M.C. Cross, D.S. Fisher
"A new theory of the spin-Peierls transition
with special relevance to the experiments on TTFCuBDT"
Phys. Rev. @B19(1979)402-419

\REF SKM SKM \Jref
Z.G. Soos, S. Kuwajima, J.E. Mihalick
"Ground-state alternation and excitation energy of $S=1/2$
linear Heisenberg antiferromagnets"
Phys. rev. @B32(1985)3124-3128

\REF AN AN \Gref
M. Aizenman, B. Nachtergaele
"Geometric Aspects of Quantum Spin States"
to appear in Commun. Math. Phys.

\REF BC BC \Bref
D. Baeriswyl, D.K. Campbell
"Interacting Electrons in Reduced Dimensions"
Plenum, New York-London 1989, NATO ASI Vol. B213

\REF DM DM \Jref
S.N. Dixit, S. Mazumdar
"Electron-electron interaction effects on
Peierls dimerization in a half-filled band"
Phys. Rev. @B29(1984)1824-1839

\REF KSSH KSSH \Jref
S. Kivelson, W.-P. Su, J.R. Schrieffer, A.J. Heeger
"Missing Bond-Charge Repulsion in the Extended Hubbard
Model: Effects in Polyacetylene"
Phys. Rev. Lett. @58(1987)1899-1902

\REF MGEW MGEW \Jref
Y. Meir, Y. Gefen, O. Entin-Wohlman
"Universal Effects of Spin-Orbit Scattering in Mesoscopic Systems"
Phys. Rev. Lett. @63(1989) 798-800

\REF EWGMO EWGMO \Jref
O. Entin-Wohlman, Y. Gefen, Y. Meir, Y. Oreg
"Effects of spin-orbit scattering in mesoscopic rings:
Canonical- versus grand-canonical-ensemble averaging"
Phys. Rev. @B45(1992) 11 890-11 895

\REF FK FK \Jref
S. Fujimoto, N. Kawakami
"Persistent currents in mesoscopic Hubbard rings
with spin-orbit interaction"
Phys. Rev. @B48(1993)17 406-17 412

\REF ML ML \Jref
D.C. Mattis, W.D. Langer
"Role of phonons and band structure in metal-insulator
phase transition"
Phys. Rev. Lett. @25(1970)376-380

\REF PVZ PVZ \Jref
J.V. Pul\'e, A. Verbeure, V.A. Zagrebnov
"Peierls-Fr\"ohlich Instability and Kohn Anomaly"
J. Stat. Phys. @76(1994)155-178

\REF SG SG \Jref
K.R. Subbaswamy, M. Grabowski
"Bond alternation, on-site Coulomb correlations,
and solitons in polyacteylene"
Phys. Rev. @B24(1981)2168-2173

\REF HM HM \Jref
G.W. Hayden, E.J. Mele
"Correlation effects and excited states
in conjugated polyners"
Phys. Rev. @B24(1986)5484-5497

\REF MR MR \Jref
E.J. Mele, M. Rice
"Semicondutor-metal transition in doped polyacetylene"
Phys. Rev. @B23(1981)5397-5412

\REF UMT UMT \Gref
K.C. Ung, S. Mazumdar, and D. Toussaint
"Metal-insulator and insulator-insulator transitions in the
quarter-filled band organic conductors"
preprint

\REF Maz Maz \Jref
S. Mazumdar
"Valence-bond approach to two-dimensional broken symmetries:
Application to $La_2CuO_4$"
Phys. Rev. @B36(1987)7190-7193

\REF TH TH \Jref
S. Tang, J.E. Hirsch
"Peierls instability in the two-dimensional half-filled
Hubbard model"
Phys. Rev. @B37(1988)9546-9558

\REF KL2 KL2 \Jref
T. Kennedy, E.H. Lieb
"An itinerant electron model with crystalline
or magnetic long range order"
Physica @138A(1986)320-358

\REF BS BS \Jref
U. Brandt, R. Schmidt
"Exact Results for the Distribution of the $f$-Level
Ground State Occupation in the Spinless Falicov-Kimball
Model"
Z. Phys. @B63(1986)45-53

\REF LM LM \Gref
J.L. Lebowitz, N. Macris
"Low-temperature phases of itinerant fermions interacting
with classical phonons: The static Holstein model"
preprint (1994)

\REF Lie1 Lie1 \Jref
E.H. Lieb
"Two Theorems on the Hubbard Model"
Phys. Rev. Lett. @62(1989) 1201-1204

\REF DLS DLS \Jref
F. J. Dyson, E.H. Lieb, B. Simon
"Phase Transitions in Quantum Spin Systems
with Isotropic and Nonisotropic Interactions"
J. Stat. Phys. @18(1978)335-383

\REF FILS FILS \Jref
J. Fr\"ohlich, R. Israel, E.H. Lieb, B. Simon
"Phase Transitions and Reflection Positivity.
I. General Theory and Long Range Lattice Models"
Commun. Math. Phys. @62(1978)1-34

\REF KLS KLS \Jref
T. Kennedy, E.H. Lieb, B.S. Shastry
"Existence of N\'eel order in some spin 1/2
Heisenberg antiferromagnets"
J. Stat. Phys. @53(1988)1019-1030

\vfill\eject

\centerline{\bf FIGURE CAPTIONS}

\bigskip\noindent
{\bf Fig. 1.} A ring containing an even number of sites,
labeled $1,\ldots,L$. The hopping matrix element $t_j$ is
associated with the bond $\{j,j+1\}$. Reflection through the plane
indicated by the dashed line interchanges the role of the hoppings
labeled $\tl$ and the ones labeled $\tr$ (see Lemma 7).

\bigskip\noindent
{\bf Fig. 2.} Graph of the function $f(t)$ defined in
\eq(f) for $a= .46 $. Recall that large $t$ corresponds to
small separation $w$ and vice versa.

\bigskip\noindent
{\bf Fig. 3.} Graph of the coefficients $x\Delta_k$ as a function
of the ratio $r=y/x$ for $L= 20 $ and $k=0,\ldots,5$. An instability
can occur when $\Delta_k(r) > \Delta_0(r)$. Note that this condition
is satisfied only in a small region of $r$ not much smaller than 1,
and only for $k=1$, which corresponds to an instability of wavelength
$L$ in real space.

\vfill\eject

\bye